\RequirePackage[l2tabu,orthodox]{nag}
\pdfoutput=1
\newif\ifarxiv
\arxivtrue
\ifarxiv
  \documentclass[format=acmsmall,timestamp=false,screen,nonacm]{acmart}
  \setcopyright{none}
\else
  \documentclass[format=acmsmall,screen,review]{acmart}
  \setcopyright{none}
  \acmPrice{}
\fi
\acmDOI{}
\acmJournal{TOMS}
  
\usepackage{amsmath}
\usepackage[subrefformat=parens,labelformat=parens,labelfont={}]{subcaption}
\usepackage{sidecap}
\usepackage{hyperref}
\usepackage{listings}
\usepackage{booktabs}
\usepackage{algorithm}
\usepackage{algpseudocode}
\algnewcommand{\LineComment}[1]{\(\triangleright\) #1}
\usepackage{
tikz,
pgfplots,
pgfplotstable,
ifthen,
}
\usetikzlibrary{fit}
\usetikzlibrary{calc}
\usetikzlibrary{patterns}
\usetikzlibrary{arrows.meta}
\tikzset{arrow/.style={line width=.1pt, arrows={-Latex[angle=30:1pt]}}}
\usepgfplotslibrary{patchplots}
\pgfplotsset{
  colormap={parula}{
    rgb255=(53,42,135)
    rgb255=(15,92,221)
    rgb255=(18,125,216)
    rgb255=(7,156,207)
    rgb255=(21,177,180)
    rgb255=(89,189,140)
    rgb255=(165,190,107)
    rgb255=(225,185,82)
    rgb255=(252,206,46)
    rgb255=(249,251,14)
  },
  colormap/parula/.style={
        colormap name=parula,
    },
}
\definecolor{viridisyellow}{HTML}{FDE725}
\definecolor{viridisyellow1}{HTML}{DCE319}
\definecolor{viridispurple}{HTML}{440154}
\definecolor{viridispurple1}{HTML}{481567}
\pgfplotsset{compat=1.18}

\newcommand{\bs}{\boldsymbol}

\newcommand{\dif}{\thinspace\mathrm{d}}

\newcommand{\grad}{\bs{\nabla}_{\!\bs{x}}}
\newcommand{\Grad}{\bs{\nabla}_{\!\bs{X}}}
\newcommand{\rgrad}{\bs{\nabla}_{\!\tilde{\bs{X}}}}

\renewcommand{\div}{\bs{\nabla}\!\cdot\!}
\newcommand{\determinant}{\text{det}}
\newcommand{\on}{\text{ on }}
\newcommand{\trace}{\text{tr}}
\newcommand{\code}[1]{\textsf{#1}}
\newcommand{\avg}[1]{\{{#1}\}}
\newcommand{\jump}[1]{[\![{#1}]\!]}
\newcommand{\minus}{\scalebox{0.75}[1.0]{\( - \)}}

\title{An abstraction for solving multi-domain problems using finite element methods}

\author{Koki Sagiyama}
\affiliation{%
  \institution{Imperial College London}
  \department{Department of Mathematics}
  \city{London}
  \country{UK}}
\email{ksagiyam@gmail.com}

\author{Lawrence Mitchell}
\affiliation{%
  \institution{Independent researcher}
  \city{Edinburgh}
  \country{UK}}
\email{lawrence@wence.uk}

\author{David A. Ham}
\affiliation{%
  \institution{Imperial College London}
  \department{Department of Mathematics}
  \city{London}
  \country{UK}}
\email{d.ham@imperial.ac.uk}

\begin{abstract}
We introduce a new abstraction for the representation and solution of multi-domain problems
using finite element methods.
This is an advance over previous work in that it achieves a single higher-level abstraction that represents multi-domain problems in the mixed variational problem formalism.
We implemented our new abstraction in UFL and Firedrake, and
validated our implementations solving
a quad-triangle mixed-cell-type problem,
a hex-quad mixed-cell-type problem, and
a fluid-structure interaction benchmark problem.

\end{abstract}

\usepackage{cleveref}
\crefname{algorithm}{Algorithm}{Algorithms}
\crefname{figure}{Fig.}{Figs.}
\crefname{table}{Table}{Tables}

\begin{document}

\maketitle

\section{Introduction}\label{S:introduction}
The coupling of physical systems defined on multiple domains
is of interest in many fields in science and engineering;
a typical example is a fluid-structure interaction problem in which
two disjoint domains,
one governed by a fluid law and the other by a structural law,
interact with each other at the interface.
Numerical simulations of such multi-domain problems are challenging, and
simulation tools that allow for a high-level specification of the problem
are crucial.

We are interested in finite element methods.
Many finite element software packages support multi-domain problems to various degrees;
these include
FEniCS \cite{fenics, Daversin-Catty2021},
FEniCSx \cite{dolfinx, Dean2023},
DUNE \cite{dune, dune_fempy},
deal.II \cite{dealii}, and
Netgen/NGSolve.
Among these, automated code generation frameworks that allow for
a high-level specification of variational problems have attracted the attention
of scientists, engineers, and mathematicians for their high level of user productivity.
The Unified Form Language (UFL) \cite{ufl} is a domain-specific language that 
allows one to express variational problems at high level and
lowers the expression to facilitate efficient low-level code generation;
UFL has been used, e.g., in FEniCS \cite{fenics}, FEniCSx \cite{dolfinx}, DUNE \cite{dune_fempy}, and Firedrake \cite{FiredrakeUserManual}.
In this context,
the work by \citet{Daversin-Catty2021}, followed by the work by \citet{Dean2023},
deserves a special attention.
They introduced a light-weight change in UFL to allow for
the expression of the multi-domain problems component-wise.
This approach avoided changes to the UFL language, but
introduced a gap between representations of 
single-domain problems and multi-domain problems;
this include necessity to express global operations
such as Gateaux derivatives component-wise at high-level.
The Python bindings for DUNE \cite{dune_fempy} take a similar approach.

One of the key advantages of UFL-based code generation finite element
frameworks is composability: variational problems, discretisations and solver
strategies can be interchanged as the user desires. The resulting PDE solver
can be combined with automated outer loop capabilities such as timestepping \cite{Kirby2025}, adjoint
simulations to solve inverse problems \cite{Farrell2013,Mitusch2019}, or deflation to
discover multiple solutions to nonlinear PDEs \cite{defcon-preprint}. Unified
abstractions underpin this composability: by representing whole classes of
mathematical structure in a consistent manner, it becomes possible for users to
substitute one for another without recoding interfaces or implementations. A
particular example of this is UFL's representation of mixed (i.e.~
multivariable) variational problems in the same way as those in a single
variable. 

This work is a step forward to achieve single higher-level abstraction that
represents multi-domain problems in the same mixed variational problem
formalism. The result is simpler user code, fewer code pathways in the
underlying framework, and hence higher productivity and reduced technical debt.
We implemented our new abstraction in UFL and Firedrake
for problems in which all domains are conforming and
of codimension 0 or 1 relative to some base domain.

\Cref{S:measures} introduces the concepts of measures and restrictions that are
essential to express multi-domain problems and \Cref{S:gateaux} introduces
representations of Gateaux derivatives in the multi-domain setting. In
\cref{S:ufl} we describe UFL abstraction that mirrors concepts introduced in
\cref{S:measures} and \cref{S:gateaux}. In \cref{S:example} we present
numerical examples to validate our implementations in UFL and Firedrake.
\Cref{S:conclusion} concludes.

\section{Measures and Restrictions}
\label{S:measures}

\input{fig_meshes}
We consider a conforming sequence of tessellations $\mathcal{T}_i$
$(i=0,\ldots,N-1)$,
where $N$ is the number of tessellations.
We denote by $\mathcal{F}_{i}^{\text{ext}}$ and $\mathcal{F}_{i}^{\text{int}}$
the sets of all exterior and interior facets of $\mathcal{T}_i$.
$\mathcal{T}_i$ $(i=1,\ldots,N-1)$ can be obtained from $\mathcal{T}_0$,
some given tesselation of a domain, either
as a subset of $\mathcal{T}_0$ or
as a subset of $\mathcal{F}_0^{\text{ext}}\cup\mathcal{F}_0^{\text{int}}$;
we call $\mathcal{T}_0$ the \emph{parent mesh} and
$\mathcal{T}_i$ $(i=1,\ldots,N-1)$ \emph{submesh}es.
For convenience, we also reflexively call $\mathcal{T}_i$ $(i=0,\ldots,N-1)$ submeshes.
We call a submesh obtained
as a subset of $\mathcal{T}_0$ a \emph{codim-0 submesh}
and one obtained
as a subset of $\mathcal{F}_0^{\text{ext}}\cup\mathcal{F}_0^{\text{int}}$ a \emph{codim-1 submesh}.
\Cref{Fi:meshes} shows an example sequence of codim-0 submeshes.
On each $\mathcal{T}_i$,
we define finite element function spaces and functions, and
we also define geometric quantities such as
coordinates and facet normals.
Note that $\mathcal{T}_0$ can be composed of cells of
different cell types, for example triangles and quadrilaterals.

\subsection{Cell integrals}\label{SS:measures_cell} \subsubsection{Cell
integration measures for multi-domain
problems}\label{SSS:measures_cell_definition}
In this section we only consider codim-0 submeshes.
We let $\dif x_i$ denote the
standard single-domain cell integration measure on $\mathcal{T}_i$. We first
consider cell integrals involving functions and geometric quantities on meshes
$\mathcal{T}_{i_k}$ ($i_k\in I$), where $I\subset\{0,1,\cdots,N-1\}$. We define
an \emph{intersection measure} $\dif x$ as:
\begin{align}
\dif x = \bigcap_{i\in I}\dif x_i.
\label{E:dx}
\end{align}
The intersection measure $\dif x$ is
a multi-domain cell integration measure
designated for cell integrations over the set intersection:
\begin{align}
\mathcal{T}=\bigcap_{i\in I}\mathcal{T}_i,
\end{align}
or any subset of $\mathcal{T}$.

\subsubsection{Examples}\label{SSS:measures_cell_example} We consider the
sequence of submeshes, $\mathcal{T}_0$, $\mathcal{T}_1$, and $\mathcal{T}_2$,
shown in \cref{Fi:meshes}. We let $V_0$, $V_1$, and $V_2$ be some scalar
function spaces on $\mathcal{T}_0$, $\mathcal{T}_1$, and $\mathcal{T}_2$, and
let $u_0$, $u_1$, and $u_2$ be some functions on $V_0$, $V_1$, and $V_2$. The
following cell integral is valid, and results in integration over the cells
marked $\text{Z}$ in \cref{Fi:meshes}:
\begin{align}
\int_{\mathcal{T}_{\text{Z}}}
(u_0u_1u_2)
\dif x_0\!\cap\!\dif x_1\!\cap\!\dif x_2.
\end{align}

\subsection{Facet integrals}\label{SS:measures_facet}
\subsubsection{Conventions}\label{SSS:measures_facet_convention}
In the following we follow the convention in
the Unified Form Language (UFL) \citep{ufl} literature and
write an integral over an exterior facet $F\in\mathcal{F}^{\text{ext}}_i$
by restricting the expression $f$ in the interior to the exterior facet $F$,
denoting it simply as $f$.
We denote the measure designated for the exterior facet integration
on the domain $\mathcal{T}_i$ as $\dif s_i$.
We denote by $\bs{n}_i$ the outward normal to the interior on $\mathcal{F}^{\text{ext}}_i$.
On the other hand, when writing an integral over an interior facet $F\in\mathcal{F}^{\text{int}}_i$,
we restrict the expression $f$ to the positive or the negative side of the interior facet $F$,
denoting it as $f^+$ or $f^-$, respectively.
We denote the measure designated for the interior facet integration
on the domain $\mathcal{T}_i$ as $\dif S_i$.
Quantities defined on $\mathcal{T}_i$ must be restricted by
either $+$ or $-$ in the $\dif S_i$-integral.
We denote by $\bs{n}_i^+$ and $\bs{n}_i^-$ the outward normals to
the $+$ and the $-$ sides on $\mathcal{F}^{\text{int}}_i$.

\subsubsection{Facet integration measures for multi-domain problems}\label{SSS:measures_facet_definition}
In this section we only consider codim-0 submeshes.
We consider facet integrals involving
functions and geometric quantities on meshes
$\mathcal{T}_{i_k}$ ($i_k\in I$),
where $I\subset\{0,1,\cdots,N-1\}$.
We define an intersection measure $\dif s$ as:
\begin{align}
\dif s=\left(
\bigcap_{i\in I^{\text{ext}}}\!\!\dif s_i
\right)\cap\left(
\bigcap_{i\in I^{\text{int}}}\!\!\dif S_i
\right),
\label{E:ds}
\end{align}
where $I^{\text{ext}}\cup I^{\text{int}}=I$ and
$I^{\text{ext}}\cap I^{\text{int}}=\emptyset$.
The intersection measure $\dif s$ is a multi-domain facet integration measure designated for facet integrations over the set intersection:
\begin{align}
\mathcal{F}=\left(
\bigcap_{i\in I^{\text{ext}}}\!\!\mathcal{F}^{\text{ext}}_i
\right)\cap\left(
\bigcap_{i\in I^{\text{int}}}\!\!\mathcal{F}^{\text{int}}_i
\right),
\end{align}
or any subset of $\mathcal{F}$. 
Functions and geometric quantities on $\mathcal{T}_i$
are to be restricted to $F\in\mathcal{F}^{\text{ext}}_i$ from the interior if $i\in I^{\text{ext}}$ and 
to $F\in\mathcal{F}^{\text{int}}_i$ either from the $+$ or the $-$ side
if $i\in I^{\text{int}}$.

Note that any facet integral can be decomposed into a sum of facet integrals so that
each integral has an intersection measure as defined in
\cref{E:ds}.

\subsubsection{Examples}\label{SSS:measures_facet_example}
We consider the sequence of submeshes,
$\mathcal{T}_0$, $\mathcal{T}_1$, and $\mathcal{T}_2$,
shown in \cref{Fi:meshes}.
We let $V_0$ and $V_1$ be some scalar DG spaces on $\mathcal{T}_0$ and $\mathcal{T}_1$ and
$V_2$ be some H(div) space on $\mathcal{T}_2$.
We then let
$u_0$, $u_1$, and $u_2$ be some functions on
$V_0$, $V_1$, and $V_2$.
The following facet integrals are valid:
\begin{align}
\int_{\mathcal{E}_{\text{A}}}
(u_0^+u_1^+(u_2\!\cdot\!\bs{n}_2))
&\dif S_0\!\cap\!\dif S_1\!\cap\!\dif s_2,\label{E:valid_F_A}\\
\int_{\mathcal{E}_{\text{B}}}
(u_0^+u_1(u_2^+\!\cdot\!\bs{n}_2^+))
&\dif S_0\!\cap\!\dif s_1\!\cap\!\dif S_2,\label{E:valid_F_B}\\
\int_{\mathcal{E}_{\text{C}}}
(u_0^+u_1)
&\dif S_0\!\cap\!\dif s_1,\label{E:valid_F_C}\\
\int_{\mathcal{E}_{\text{I}}}
(u_0^+u_1(u_2\!\cdot\!\bs{n}_2))
&\dif S_0\!\cap\!\dif s_1\!\cap\!\dif s_2,\label{E:valid_F_I}\\
\int_{\mathcal{E}_{\text{A}}}
(u_0^+u_1^+(u_2\!\cdot\!\bs{n}_2))
\dif S_0\!\cap\!\dif S_1\!\cap\!\dif s_2
&+\int_{\mathcal{E}_{\text{I}}}
(u_0^+u_1(u_2\!\cdot\!\bs{n}_2))
\dif S_0\!\cap\!\dif s_1\!\cap\!\dif s_2.\label{E:valid_F_A_F_I}
\end{align}

\subsection{Mixed cell-facet integrals}\label{SS:measures_mixed}

\subsubsection{Mixed cell-facet integration measures for multi-domain problems}\label{SSS:measures_mixed_definition}
In this section we consider a more general case in which both codim-0 and codim-1 submeshes are involved.
We consider integrals involving
functions and geometric quantities on meshes
$\mathcal{T}_{i_k}$ ($i_k\in I$),
where $I\subset\{0,1,\cdots,N-1\}$.
We define an intersection measure $\dif z$ as:
\begin{align}
\dif z=\left(
\bigcap_{i\in I^{\text{cell}}}\dif x_i
\right)\cap\left(
\bigcap_{i\in I^{\text{ext}}}\!\!\dif s_i
\right)\cap\left(
\bigcap_{i\in I^{\text{int}}}\!\!\dif S_i
\right),
\label{E:dz}
\end{align}
where
$I^{\text{cell}}\cup I^{\text{ext}}\cup I^{\text{int}}=I$
and
$I^{\text{cell}}\cap I^{\text{ext}}\cap I^{\text{int}}=\emptyset$;
$\mathcal{T}_i$ $(i\in I^{\text{cell}})$ are codim-1 submeshes
and
$\mathcal{T}_i$ $(i\in I^{\text{ext}}\cup I^{\text{int}})$ are codim-0 submeshes.
The intersection measure $\dif z$ is a multi-domain integration measure designated for integrations over the set intersection:
\begin{align}
\mathcal{E}=
\left(
\bigcap_{i\in I^{\text{cell}}}\!\!\mathcal{T}_i
\right)\cap\left(
\bigcap_{i\in I^{\text{ext}}}\!\!\mathcal{F}^{\text{ext}}_i
\right)\cap\left(
\bigcap_{i\in I^{\text{int}}}\!\!\mathcal{F}^{\text{int}}_i
\right),
\end{align}
or any subset of $\mathcal{E}$.

\subsubsection{Examples}\label{SSS:measures_mixed_example}
We consider a sequence of submeshes,
$\mathcal{T}_0$, $\mathcal{T}_1$, and $\mathcal{T}_2$,
where
$\mathcal{T}_0$ and $\mathcal{T}_1$ are codim-0 submeshes and
$\mathcal{T}_2$ is a codim-1 submesh.
Suppose that
$I=\{0,1,2\}$, and
$I^{\text{int}}=\{0\}$,
$I^{\text{ext}}=\{1\}$, and
$I^{\text{cell}}=\{2\}$.
We let $V_0$, $V_1$, and $V_2$ be
some H(curl) space on $\mathcal{T}_0$,
some H(div) space on $\mathcal{T}_1$, and
some vector DG space on $\mathcal{T}_2$, respectively.
We then let
$u_0$, $u_1$, and $u_2$ be some functions on
$V_0$, $V_1$, and $V_2$.
The following integral is valid:
\begin{align}
\int_{\mathcal{E}}
(
(u_0^+\times\bs{n}_0^++u_0^-\times\bs{n}_0^-)\!\cdot\!u_2+
u_1\!\cdot\!\bs{n}_1
)
\dif S_0\!\cap\!\dif s_1\!\cap\!\dif x_2,\label{E:valid_E}.
\end{align}

\section{Gateaux derivatives}\label{S:gateaux}
In this section we consider
the residual and the Jacobian
of a multi-domain variational problem.
We write them both in component-form and in compact, monolithic form.
As seen in \cref{S:ufl}, this highlights the difference
between UFL abstraction introduced in \citet{Daversin-Catty2021} and
that that we implemented in this work.

Suppose that the solution space is given as the following:
\begin{align}
V = V_{i_0}\times V_{i_1}\times\cdots\times V_{i_{M-1}},
\label{E:V_comp}
\end{align}
where
$V_{i_m}$ is a function space defined on $\mathcal{T}_{i_m}$, and
$M$ is the number of components.
We consider a system of PDEs that is solved for:
\begin{align}
(u_{i_0},...,u_{i_{M-1}})\in V_{i_0}\times V_{i_1}\times\cdots\times V_{i_{M-1}},
\label{E:u_comp}
\end{align}
or, written monolithically:
\begin{align}
u\in V.
\label{E:u}
\end{align}
The test function, provided that it is defined to lie in the same space,
is similarly written as:
\begin{align}
(v_{i_0},...,v_{i_{M-1}})\in V_{i_0}\times V_{i_1}\times\cdots\times V_{i_{M-1}},
\label{E:v_comp}
\end{align}
or monolithically as:
\begin{align}
v\in V.
\label{E:v}
\end{align}
The residual of the system is written as:
\begin{align}
F(u_{i_0},...,u_{i_{M-1}};v_{i_0},...,v_{i_{M-1}}),
\label{E:F(u;v)_comp}
\end{align}
or monolithically as:
\begin{align}
F(u;v).
\label{E:F(u;v)}
\end{align}
Iterative solution techniques such as Newton's method require
taking the Gateaux derivative of the residual $F$ to obtain the Jacobian
$J$.
In component form $J$ is written as:
\begin{align}
J(u_{i_0},...,u_{i_{M-1}};v_{i_0},...,v_{i_{M-1}},\delta u_{i_0},...,\delta u_{i_{M-1}})
=\sum\limits_{i_m=0,...,M-1}\delta_{u_{i_m}}F(u_{i_0},...,u_{i_{M-1}};v_{i_0},...,v_{i_{M-1}}),
\label{E:J(u;v_du)_comp}
\end{align}
where $\delta_{u_{i_m}}(\cdot)$ represents the Gateaux derivative
with respect to the solution component $u_{i_m}$, and
$\delta u_{i_m}$ is the direction of perturbation.
Alternatively, one can write $J$ monolithically as:
\begin{align}
J(u;v,\delta u)=\delta_uF(u;v),
\label{E:J(u;v_du)}
\end{align}
to obtain a multi-variable system that is identical in the form to the single-variable system, where
$\delta_{u}(\cdot)$ represents the Gateaux derivative
with respect to the total solution $u$, and
$\delta u$ is the direction of perturbation.

We note that, if we assemble $J$,
the $(r,c)$ off-diagonal block of $J$
is to represent coupling between $u_{i_r}$ and $u_{i_c}$.

\section{Representation in the Unified Form Language}\label{S:ufl}
The Unified Form Language (UFL) \citep{ufl} is a domain-specific language
that allows for the expression of variational formulations symbolically, and provides algorithms that lower these formulations to a form that is suitable for low-level code generation.

In this section we discuss changes that we made to support multi-domain
problems seamlessly in UFL. For simplicity, in the below, we assume that the
standard Galerkin finite element formulation is used, though this is not a
limitation of UFL or of the extensions presented here.

\begin{figure}
\centering
\begin{subfigure}[b]{0.3\textwidth}
\centering
\begin{tikzpicture}
\tikzstyle{fs}=[draw=black, ultra thick, rounded corners=20pt, minimum size=10pt, inner sep=2pt]
\tikzstyle{mesh}=[draw=black, ultra thick, densely dashed, rounded corners=20pt, minimum size=10pt, inner sep=2pt]
\tikzstyle{elem}=[draw=black, ultra thick, loosely dashed, rounded corners=20pt, minimum size=10pt, inner sep=2pt]
\draw [fs] (0,0) rectangle (4.5,-6);
\draw [mesh] (0.25,-0.5) rectangle (4.25,-3.75);
\draw [elem] (0.25,-2.25) rectangle (4.25,-5.5);
\node at (2.25, 0.5) [font=\sffamily, anchor=center] {FunctionSpace};
\node at (2.25,-1.5) [font=\sffamily, anchor=center] {AbstractDomain};
\node at (2.25,-3) [font=\sffamily, anchor=center] {AbstractCell};
\node at (2.25,-4.5) [font=\sffamily, anchor=center] {AbstractFiniteElement};
\end{tikzpicture}
\caption{Abstract}
\label{Fi:abstraction_0}
\end{subfigure}
\hfill
\begin{subfigure}[b]{0.3\textwidth}
\centering
\begin{tikzpicture}
\tikzstyle{fs}=[draw=black, ultra thick, rounded corners=20pt, minimum size=10pt, inner sep=2pt]
\tikzstyle{mesh}=[draw=black, ultra thick, densely dashed, rounded corners=20pt, minimum size=10pt, inner sep=2pt]
\tikzstyle{elem}=[draw=black, ultra thick, loosely dashed, rounded corners=20pt, minimum size=10pt, inner sep=2pt]
\draw [fs] (0,0) rectangle (4.5,-6);
\draw [mesh] (0.25,-0.5) rectangle (4.25,-3.75);
\draw [elem] (0.25,-2.25) rectangle (4.25,-5.5);
\node at (2.25, 0.5) [font=\sffamily, anchor=center] {FunctionSpace};
\node at (2.25,-1.5) [font=\sffamily, anchor=center] {Mesh};
\node at (2.25,-3) [font=\sffamily, anchor=center] {Cell};
\node at (2.25,-4.5) [font=\sffamily, anchor=center] {FiniteElement};
\end{tikzpicture}
\caption{Single-domain}
\label{Fi:abstraction_1}
\end{subfigure}
\hfill
\begin{subfigure}[b]{0.3\textwidth}
\centering
\begin{tikzpicture}
\tikzstyle{fs}=[draw=black, ultra thick, rounded corners=20pt, minimum size=10pt, inner sep=2pt]
\tikzstyle{mesh}=[draw=black, ultra thick, densely dashed, rounded corners=20pt, minimum size=10pt, inner sep=2pt]
\tikzstyle{elem}=[draw=black, ultra thick, loosely dashed, rounded corners=20pt, minimum size=10pt, inner sep=2pt]
\draw [fs] (0,0) rectangle (4.5,-6);
\draw [mesh] (0.25,-0.5) rectangle (4.25,-3.75);
\draw [elem] (0.25,-2.25) rectangle (4.25,-5.5);
\node at (2.25, 0.5) [font=\sffamily, anchor=center] {FunctionSpace};
\node at (2.25,-1.5) [font=\sffamily, anchor=center] {MeshSequence};
\node at (2.25,-3) [font=\sffamily, anchor=center] {CellSequence};
\node at (2.25,-4.5) [font=\sffamily, anchor=center] {MixedElement};
\end{tikzpicture}
\caption{Multi-domain}
\label{Fi:abstraction_2}
\end{subfigure}
\caption{
UFL \code{FunctionSpace} structures.
\subref{Fi:abstraction_0} \code{FunctionSpace} abstraction.
\subref{Fi:abstraction_1} Conventional \code{FunctionSpace} constituents for a single-domain problem.
\subref{Fi:abstraction_2} \code{FunctionSpace} constituents for a multi-domain problem.
}
\label{fig:three graphs}
\end{figure}
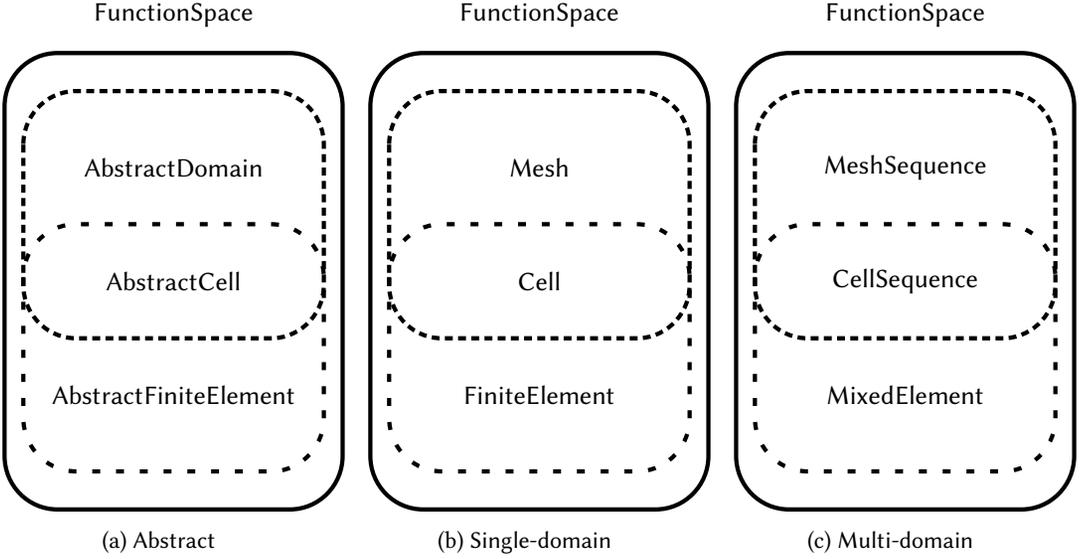

\subsection{Single-domain problems}\label{SS:ufl_single}
In a conventional single-domain problem,
a \code{FunctionSpace},
e.g., one representing the solution space,
is defined by
a \code{Mesh} and a \code{FiniteElement},
which are defined on a common \code{Cell};
\cref{Fi:abstraction_1} depicts the \code{FunctionSpace} structure for a single-domain problem.
Then, on a \code{FunctionSpace}, one can for instance define
\code{Coefficient}s\footnote{\code{Coefficient} is the UFL notation. Firedrake's \code{Coefficient} subclass is called \code{Function}}, i.e., functions, and
\code{Argument}s, i.e., unknown functions.
One can also define \code{GeometricQuantity}s on the \code{Mesh}
representing, e.g.,
the coordinates and the outward facet normals.
We express the solution of the problem
as a \code{Coefficient} defined on the \code{FunctionSpace}
representing the solution space
and the test function
as an \code{Argument}, or, more specifically, as a \code{TestFunction},
defined on the same \code{FunctionSpace}.
The solution, test function, other \code{Coefficient}s, \code{GeometricQuantity}s, and operators, such as
\code{grad} representing the gradient and
\code{inner} representing the inner product,
are then used to express an integral expression
along with a \code{Measure} that specifies the single integration domain and the integration type on that domain.
The residual of the problem is expressed as a sum of the integral expressions.
The corresponding Jacobian is expressed using the global operator \code{derivative},
taking the Gateaux derivative of the residual with respect to the solution.

\Cref{Code:ufl_example_single_domain} shows
a typical UFL code to express the Jacobian
of a Poisson problem on a single domain $\Omega$ representing:
\begin{align}
\int_{\Omega}
\nabla u\!\cdot\!\nabla\delta u\dif V,
\label{E:ufl_example_single_domain}
\end{align}
where 
$u\in V$ represents the solution and
$\delta u\in V$ represents the test function,
where $V$ represents the P1 function space on $\Omega$.
Note that the function spaces in UFL expressions are unconstrained;
boundary conditions are applied by the problem solving environment
such as Firedrake \citep{FiredrakeUserManual}
where necessary.

\begin{lstlisting}[
language=Python,
caption={
UFL code example: Jacobian for the Poisson problem (single-domain).
},
captionpos=b,
label=Code:ufl_example_single_domain,
]
from ufl import *
dim = 2
coord_degree = 1
cell = Cell("triangle")
mesh = Mesh(LagrangeElement(cell, coord_degree, (dim,)))
elem = LagrangeElement(cell, 1, ())  # P1
dx = Measure("dx", mesh)
V = FunctionSpace(mesh, elem)
u = Coefficient(V)
du = TestFunction(V)
F = inner(grad(u), grad(du)) * dx
J = derivative(F, u)
\end{lstlisting}

\subsection{Multi-domain problems}\label{SS:ufl_multi}
\code{Mesh}, \code{FiniteElement}, and \code{Cell}
are child classes of
\code{AbstractDomain}, \code{AbstractFiniteElement}, and \code{AbstractCell},
respectively;
\cref{Fi:abstraction_0} depicts this abstraction.
To seamlessly integrate multi-domain problems respecting
the existing \code{FunctionSpace} abstraction in UFL,
we made changes so that a single \code{FunctionSpace}
can represent a product space composed of function spaces on different meshes.
This highlights the difference from \citet{Daversin-Catty2021}, in which
a product space was represented as a \code{MixedFunctionSpace},
which was merely a tuple of component \code{FunctionSpace}s;
this difference is illustrated by the right-hand side of \cref{E:V_comp} and the left-hand side of \cref{E:V_comp}.
Specifically, we allow the definition of a \code{FunctionSpace} with
a \code{MeshSequence} and a \code{MixedElement},
which share a \code{CellSequence};
\code{MeshSequence}, \code{MixedElement}, and \code{CellSequence}
are, respectively, child classes of
\code{AbstractDomain}, \code{AbstractFiniteElement}, and \code{AbstractCell}.
\Cref{Fi:abstraction_2} depicts our \code{FunctionSpace} structure for a multi-domain problem.
\code{MeshSequence} is a new class for representing a sequence of meshes
appearing in the product space.
\code{MixedElement} previously existed to represent a sequence of finite elements
appearing in the product space,
but we generalised it to support finite elements of inhomogeneous cell types.
\code{CellSequence} is a new class for representing a sequence of potentially inhomogeneous cells
appearing in the product space;
i.e., a \code{CellSequence} represents the common cell type
for the \code{MeshSequence} and the \code{MixedElement}
that define the product space.
One can then define, e.g., 
\code{Coefficient}s and \code{Argument}s,
directly on the \code{FunctionSpace} representing the product space,
and use \code{Indexed} class that has existed to represent their components as needed that
keeps the reference to the original unbroken \code{Coefficient}s and \code{Argument}s
defined on the \code{MeshSequence}.
This is in contrast with having to construct the component
\code{Coefficient}s or \code{Argument}s on the component spaces explicitly;
compare \cref{E:u_comp} and \cref{E:v_comp} with \cref{E:u} and \cref{E:v}.
An integrand can then be expressed using local operators as normal.
To express an integral in a multi-domain problem,
we let \code{Measure} take an additional arguement, \code{intersect\_measures},
to represent the intersection measures such as
\cref{E:dx}, \cref{E:ds}, and \cref{E:dz} introduced in \Cref{S:measures};
one of the domains, codim-0 or codim-1, is chosen as the \emph{primal} integration domain
that is to define the iteration set used by the assembler.
The residual can then be expressed as a sum of integrals, and
the Jacobian can be expressed by taking \code{derivative}
of the residual directly with respect to the solution;
in other words,
one can express Gateaux derivatives
without explicitly working with the
component \code{Coefficient}s representing the solution and the
component \code{Argument}s representing the test function.
This is better understood by comparing
\cref{E:F(u;v)_comp} and \cref{E:J(u;v_du)_comp} with
\cref{E:F(u;v)} and \cref{E:J(u;v_du)}.
This simplification is directly reflected by simplification in
expressing and lowering the Gateaux derivatives
and other global operators.

\Cref{Code:ufl_example_multi_domain} shows the UFL code to express the Jacobian of a Poisson’s problem solved on a mesh composed of quadrilaterals and triangles.
We extract quadrilateral part of the mesh $\Omega^q$ and triangular part of the mesh $\Omega^t$ as submeshes,
and denote the interface by $\Gamma^{\text{interf}}$.
We define Q1 and P1 function spaces $V^q$ and $V^t$, respectively on
$\Omega^q$ and $\Omega^t$, and
define the product space $V=V^q\times V^t$.
We then define the solution $u=(u^q,u^t)\in V$ and
the test function $\delta u=(\delta u^q, \delta u^t)\in V$.
Using the classical interior penalty method to \emph{glue} the solution at the interface,
the residual can be written as:
\begin{align}
&\int_{\Omega^q}
\nabla u^q\!\cdot\!\nabla\delta u^q\dif V+
\int_{\Omega^t}
\nabla u^t\!\cdot\!\nabla\delta u^t\dif V\notag\\
&-\int_{\Gamma^{\text{interf}}}\avg{\nabla u}\!\cdot\!\jump{\delta u}\dif A
-\int_{\Gamma^{\text{interf}}}\jump{u}\!\cdot\!\avg{\nabla{\delta u}}\dif A
+\frac{C}{h}\int_{\Gamma^{\text{interf}}}\jump{u}\jump{\delta u}\dif A,
\label{E:ufl_example_multi_domain}
\end{align}
where $C$ is constant, $h$ is the element size, and: 
\begin{align}
\avg{\nabla u}&=(\nabla u^q+\nabla u^t)/2,\label{E:avg}\\
\jump{u}&=u^q\bs{n}^q+u^t\bs{n}^t\label{E:jump},
\end{align}
where $\bs{n}^q$ and $\bs{n}^t$ are unit outward normals to $\Omega^q$ and $\Omega^t$.
The corresponding Jacobian can be expressed as the Gateaux derivative of the residual
with respect to the total solution $u$.
\begin{lstlisting}[
language=Python,
caption={
UFL code example: Jacobian for the Poisson problem (multi-domain).
},
captionpos=b,
label=Code:ufl_example_multi_domain,
]
dim = 2
coord_degree = 1
cell_q = Cell("quadrilateral")
cell_t = Cell("triangle")
mesh_q = Mesh(LagrangeElement(cell_q, coord_degree, (dim,)))
mesh_t = Mesh(LagrangeElement(cell_t, coord_degree, (dim,)))
mesh = MeshSequence([mesh_q, mesh_t])
assert mesh.ufl_cell() == CellSequence([cell_q, cell_t])
elem_q = LagrangeElement(cell_q, 1, ())  # Q1
elem_t = LagrangeElement(cell_t, 1, ())  # P1
elem = MixedElement([elem_q, elem_t], make_cell_sequence=True)
assert elem.cell == CellSequence([cell_q, cell_t])
dx_q = Measure("dx", mesh_q)
dx_t = Measure("dx", mesh_t)
ds_q = Measure(
    "ds", mesh_q,
    intersect_measures=(Measure("ds", mesh_t),),
)
ds_t = Measure(
    "ds", mesh_t,
    intersect_measures=(Measure("ds", mesh_q),),
)
V = FunctionSpace(mesh, elem)
u = Coefficient(V)
v = TestFunction(V)
u_q, u_t = split(u)
v_q, v_t = split(v)
n_q = FacetNormal(mesh_q)
n_t = FacetNormal(mesh_t)
C = 100.
h = 0.1                # mesh size
interface_id = 999     # subdomain_id for the interface
F = (
    inner(grad(u_q), grad(v_q)) * dx_q +
    inner(grad(u_t), grad(v_t)) * dx_t
    -inner(
        (grad(u_q) + grad(u_t)) / 2,
        (v_q * n_q + v_t * n_t)
    ) * ds_q(interface_id)
    -inner(
        (u_q * n_q + u_t * n_t),
        (grad(v_q) + grad(v_t)) / 2
    ) * ds_t(interface_id)
    + C / h * inner(u_q - u_t, v_q - v_t) * ds_q(interface_id)
)
J = derivative(F, u)
\end{lstlisting}

\subsection{Firedrake}\label{SS:ufl_firedrake}
Firedrake \citep{FiredrakeUserManual} is an automated system for the solution of partial differential equations using the finite element method.
Firedrake composes multiple packages, including UFL,
and achieve separation of concerns.
Although the goal of this paper is to introduce
an abstraction for solving multi-domain problems,
we summarise in the below the other changes that we made
to support multi-domain problems in the Firedrake ecosystem.

Firedrake uses the DMPlex \citep{KnepleyKarpeev09,LangeMitchellKnepleyGorman2015,LangeKnepleyGorman2015} component of PETSc,
the Portable, Extensible Toolkit for Scientific Computation
\citep{petsc-web-page,petsc-user-ref,petsc-efficient},
to represent and manage unstructured meshes.
Submeshes are constructed with DMPlex
from a parent mesh that has already been
distributed over the parallel processes
by extracting marked entities, e.g., cells and facets.
We note that, though our new feature for multi-domain problems naturally inherit Firedrake's prallelism,
submeshes are partitioned using the partition of the parent mesh
regardless of the sizes of the submeshes;
this could potentially cause load imbalance, and is to be improved in the future work.

Each entity in a mesh/submesh is identified by the entity ID, and
the submesh constructor induces an \emph{entity-entity map}
that associate the entity IDs on the submesh with
those on the parent mesh.
When submesh construction is nested,
we can compose entity-entity maps to associate generated submeshes.

As explained in detail in \Cref{S:ufl},
we then express the multi-domain variational problem
on those submeshes and
lowers the expressions using UFL.
TSFC,
the two-stage form compiler for Firedrake \cite{tsfc},
then takes the low-level UFL symbolic expressions,
rewrites them as tensor algebra expressions, and
generates efficient codes for the element-local assembly;
\cref{alg:tsfc_kernel} shows the element-local kernel signature.
In \cref{alg:tsfc_kernel} 
$t$ is the element-local result tensor,
$w_0, \dots$ are active coefficients, and
$g_0, \dots$ are mesh arguments such as entity orientation.
In general mesh arguments of a certain kind, e.g., entity orientation,
are required for multiple submeshes
in multi-domain problems.

\begin{algorithm}
\caption{TSFC element-local kernel}\label{alg:tsfc_kernel}
\begin{algorithmic}
\Function{element\_local\_kernel}{$t, w_0, \dots, g_0, \dots$}
\State\LineComment{Assemble the element-local result tensor $t$}
\State $t \gets \dots$
\EndFunction
\end{algorithmic}
\end{algorithm}

Firedrake's PyOP2 \cite{pyop2},
the framework for parallel computations on unstructured meshes,
then wraps the element-local kernel to assemble the relevant global tensors
looping over entities of, a given submesh.
\cref{alg:pyop2_kernel} shows the global kernel.
In \cref{alg:pyop2_kernel},
$T$ is the global output tensor,
$W_0, \dots$ are the global input tensors,
$G_0, \dots$ are the global mesh arguments for, e.g., entity orientations, and
$\text{map}_0, \dots$ are entity-entity maps between relevant submeshes.
$\text{map}_S, S\in\{T, W_0, \dots, G_0, \dots\}$, is the map
from the entity ID, $e_S$, on the mesh on which $S$ is defined
to the packing/unpacking indices, and
$\text{cmap}_S$ is the composed map that maps $e_X$ to $e_S$.

\begin{algorithm}
\caption{PyOP2 global kernel}\label{alg:pyop2_kernel}
\begin{algorithmic}
\Function{global\_kernel}{$T, W_0, \dots, G_0, \dots, \text{map}_T, \text{map}_{W_0}, \dots, \text{map}_{G_0}, \dots, \text{map}_0, \dots$}
\For{\texttt{$e_X$ in entity IDs on submeshX}}
\State\LineComment{Initialise the element-local result tensor $t$}
\State $t \gets 0$
\State\LineComment{Pack global input tensors into element-local tensors}
\State $\text{cmap}_{W_0} \gets \text{ compose select maps in }\{\text{map}_0, \dots\}$
\State $w_0 \gets W_0[\text{map}_{W_0}[\text{cmap}_{W_0}[e_X]]]$
\State $\dots$
\State $\text{cmap}_{G_0} \gets \text{ compose select maps in }\{\text{map}_0, \dots\}$
\State $g_0 \gets G_0[\text{map}_{G_0}[\text{cmap}_{G_0}[e_X]]]$
\State $\dots$
\State\LineComment{Execute local kernel, and assemble $t$}
\State \Call{element\_local\_kernel}{$t, w_0, \dots, g_0, \dots$}
\State\LineComment{Unpack $t$ into the global output tensor $T$}
\State $\text{cmap}_T \gets \text{ compose select maps in }\{\text{map}_0, \dots\}$
\State $T[\text{map}_T[\text{cmap}_T[e_X]]] \gets t$
\EndFor
\EndFunction
\end{algorithmic}
\end{algorithm}

\section{Numerical examples}\label{S:example}
In this section we validate our implementations
in Firedrake and UFL
solving three problems:
a quad-triangle mixed cell-type problem \cref{SS:example_mixed},
a hex-quad mixed cell-type problem \cref{SS:example_codim1}, and
a fluid-structure interaction problem \cref{SS:example_fsi}.
In these problems, we used
the Portable, Extensible Toolkit for Scientific Computation (PETSc)
\citep{petsc-web-page, petsc-user-ref, petsc-efficient}
for linear and nonlinear solvers.

\subsection{Quad-triangle mixed cell-type problem}\label{SS:example_mixed}

\begin{figure}
\centering
\begin{tikzpicture}
\begin{axis}[
             xmin= -0.75, xmax= 1.75,
             ymin= -0.25, ymax= 1.5,
             point meta min= 0., point meta max= 1.0,
             axis equal image,
             colormap/viridis,
             width=300pt,
             height=150pt,
             xtick={0.001, 1.},
             xticklabels={0., 1.},
             ytick={0.001, 1.},
             yticklabels={0., 1.},
             axis line style={draw=none},
             axis x line=middle,
             axis y line=middle,
             tick style={draw=none},
            ]
\addplot[patch,
          patch type=rectangle,
          point meta=explicit,
          shader=faceted interp, 
          opacity=1.,
         ] table[x=x, y=y, meta=mesh_quad] {mixed_cell_mesh_quad.dat};
\addplot[patch,
          patch type=triangle,
          point meta=explicit,
          shader=faceted interp, 
          opacity=1.,
         ] table[x=x, y=y, meta=mesh_tri] {mixed_cell_mesh_tri.dat};
\node[] (start) at (0.60, 0.55) {};
\node[] (end0) at (0.45, 0.30) {};
\node[] (end1) at (0.77, 0.42) {};
\node[] (anchor=west) at ($(start)$) {$\Gamma^{\text{interf}}$};
\path[draw=black,-latex,shorten <= 5pt,shorten >= 2pt] ($(start)$)--($(end0)$);
\path[draw=black,-latex,shorten <= 5pt,shorten >= 2pt] ($(start)$)--($(end1)$);
\draw[-latex,draw=black,shorten <= 2pt,shorten >= 2pt] (1,0) -- (1.25,0);
\node at(axis cs:1.25,0) [anchor=north]{$x$};
\draw[-latex,draw=black,shorten <= 2pt,shorten >= 2pt] (0,1) -- (0,1.25);
\node at(axis cs:0,1.25) [anchor=east]{$y$};
\addplot[patch,
         patch type=rectangle,
         point meta=explicit,
         shader=faceted interp,
         opacity=1.,
        ] table[x=x, y=y, meta=c, row sep=crcr] {
x y c\\
1.4 0.0 1.\\
1.5 0.0 1.\\
1.5 0.1 1.\\
1.4 0.1 1.\\
};
\addplot[patch,
         patch type=triangle,
         point meta=explicit,
         shader=faceted interp,
         opacity=1.,
        ] table[x=x, y=y, meta=c, row sep=crcr] {
x y c\\
1.4 0.15 0.\\
1.5 0.15 0.\\
1.4 0.25 0.\\
};
\node[] (anchor=south east) at (1.625, 0.20) {$\Omega^t$};
\node[] (anchor=south east) at (1.625, 0.05) {$\Omega^q$};
\end{axis}
\end{tikzpicture}
\caption{
Quad-triangle mixed cell-type problem setup.
}
\label{Fi:example_mixed}
\end{figure}
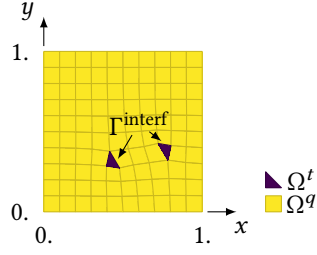

\begin{table}
\centering
\caption{Quad-triangle mixed cell-type problem: parameters.\label{Ta:example_mixed_parameters}}
\begin{tabular}{ clc } 
 \toprule
Parameter & Description & Value \\ 
\midrule
$C$ & Interior penalty constant &$100.$ \\
$n$ & Mesh refinement level & $\{0,1,2,3\}$ \\
$h$ & Mesh size & $0.10/2^n$ \\
$p$ & Degree of polynomial & $\{1,2,3,4\}$ \\
\bottomrule
\end{tabular}
\end{table}

\begin{table}
\centering
\caption{Quad-triangle mixed cell-type problem: solutions, test functions, and function spaces.\label{Ta:example_mixed_functions}}
\begin{tabular}{ ccccl } 
 \toprule
Solution & Test function & family & degree & Description \\ 
\midrule
$u^q\in V^q$ & $\delta u^q\in V^q_0$ & Q & $p$ & Field in $\Omega^q$\\
$u^t\in V^t$ & $\delta u^t\in V^t_0$ & P & $p$ & Field in $\Omega^t$\\
\bottomrule
\end{tabular}
\end{table}

We solved a quad-triangle mixed cell-type problem
introduced in \Cref{S:ufl}
now using a concrete domain, external forces, and boundary conditions,
and performed a convergence study.

\Cref{Fi:example_mixed} shows the unit square domain
that is composed of two disjoint domains $\Omega^q$ and $\Omega^t$.
We denote by $\Gamma^{\text{interf}}$ the interface between the two domains.

We solve the Poisson problem in $\Omega^q$ and $\Omega^t$
with boundary conditions on
$\partial\Omega^q\setminus\Gamma^{\text{interf}}$ and
$\partial\Omega^t\setminus\Gamma^{\text{interf}}$,
gluing solutions on $\Gamma^{\text{interf}}$
using the classical interior penalty method.
Specifically, we solve the following:

$  $\\
Find $(u^q,u^t)\in V^q\times V^t$ such that:
\begin{align}
&\int_{\Omega^q}
\nabla u^q\!\cdot\!\nabla\delta u^q\dif V+
\int_{\Omega^t}
\nabla u^t\!\cdot\!\nabla\delta u^t\dif V\notag\\
&-\int_{\Gamma^{\text{interf}}}\avg{\nabla u}\!\cdot\!\jump{\delta u}\dif A
-\int_{\Gamma^{\text{interf}}}\jump{u}\!\cdot\!\avg{\nabla{\delta u}}\dif A
+\frac{C}{h}\int_{\Gamma^{\text{interf}}}\jump{u}\jump{\delta u}\dif A\notag\\
&=
\int_{\Omega^q}f\delta u^q\dif V +
\int_{\Omega^t}f\delta u^t\dif V
\quad
\forall(\delta u^q, \delta u^t)\in V^q_0\times V^t_0,\\
u^q&=u_e\quad\text{on }\partial\Omega^q\setminus\Gamma^{\text{interf}},\\
u^t&=u_e\quad\text{on }\partial\Omega^t\setminus\Gamma^{\text{interf}},
\label{E:example_mixed}
\end{align}
where $\avg{\cdot}$ and $\jump{\cdot}$ operators are as defined in \cref{E:avg} and \cref{E:jump}, and:
\begin{align}
u_e=\cos(2\pi x)\cos(2\pi y),
\end{align}
and:
\begin{align}
f=8\pi^2u_e.
\end{align}
\Cref{Ta:example_mixed_parameters} summarises the parameters and
\Cref{Ta:example_mixed_functions} summarises solutions, test functions, and function spaces on which they are defined.
This is a manufactured problem whose solution is known to be $u_e$.

\cref{Fi:example_mixed} shows
a quadrilateral mesh of $\Omega^q$ and
a triangular mesh of $\Omega^t$.
We solved the Poisson problem
for each refinement level $n\in\{0,1,2,3\}$
for each polynomial degree $p\in\{1,2,3,4\}$,
and performed a convergence study by computing the $L^2$- and $H^1$-norm errors of the solutions defined, respectively, as:
\begin{subequations}
\begin{align}
&(\|u^q-u_e\|_{L^2(\Omega^q)}^2+\|u^t-u_e\|_{L^2(\Omega^t)}^2)^{1/2},\\
&(\|u^q-u_e\|_{H^1(\Omega^q)}^2+\|u^t-u_e\|_{H^1(\Omega^t)}^2)^{1/2}.
\end{align}
\end{subequations}
Refinement level $n$ corresponds to the mesh
obtained by uniformly refining the mesh shown in
\cref{Fi:example_mixed} $n$ times.
We used MUMPS parallel sparse direct solver \citep{mumps1, mumps2} via PETSc to solve the linear system.
\Cref{Ta:example_mixed_L2error,Ta:example_mixed_H1error}
summarise the $\log_2$ of $L^2$- and $H^1$-norm errors, 
where we observe the expected convergence behaviour.

\begin{table}
\centering
\caption{Mixed cell-type problem: $\log_2$ of $L^2$-norm errors for polynomial degrees $p\in\{1,2,3,4\}$ and refinement levels $n\in\{0,1,2,3\}$.\label{Ta:example_mixed_L2error}}
\begin{tabular}{ crrrrrrrr } 
 \toprule
      & $p=1$\phantom{0} &\textbf{rate}& $p=2$\phantom{0} &\textbf{rate}& $p=3$\phantom{0} &\textbf{rate}& $p=4$\phantom{0} &\textbf{rate}\\ 
\midrule
$n=0$ & \phantom{0}-5.0869 &             & \phantom{0}-9.8884 &             & -14.5569 &             & -19.4193 &\\
$n=1$ & \phantom{0}-7.0780 &\textbf{1.99}&           -12.8796 &\textbf{2.99}& -18.5426 &\textbf{3.99}& -24.4162 &\textbf{5.00}\\
$n=2$ & \phantom{0}-9.0756 &\textbf{2.00}&           -15.8768 &\textbf{3.00}& -22.5395 &\textbf{4.00}& -29.4137 &\textbf{5.00}\\
$n=3$ &           -11.0749 &\textbf{2.00}&           -18.8758 &\textbf{3.00}& -26.5388 &\textbf{4.00}& -34.4124 &\textbf{5.00}\\
\bottomrule
\end{tabular}
\end{table}
\begin{table}
  \centering
\caption{Mixed cell-type problem: $\log_2$ of $H^1$-norm errors for polynomial degrees $p\in\{1,2,3,4\}$ and refinement levels $n\in\{0,1,2,3\}$.\label{Ta:example_mixed_H1error}}
\begin{tabular}{ crrrrrrrr } 
 \toprule
      & $p=1$\phantom{0} &\textbf{rate}& $p=2$\phantom{0} &\textbf{rate}& $p=3$\phantom{0} &\textbf{rate}& $p=4$\phantom{0} &\textbf{rate}\\ 
\midrule
$n=0$ & \phantom{0}-0.2728 &             & \phantom{0}-3.8664 &             & \phantom{0}-8.0048 &             & -12.4932 &\\
$n=1$ & \phantom{0}-1.2752 &\textbf{1.00}& \phantom{0}-5.8635 &\textbf{2.00}&           -10.9965 &\textbf{2.99}& -16.4946 &\textbf{4.00}\\
$n=2$ & \phantom{0}-2.2754 &\textbf{1.00}& \phantom{0}-7.8621 &\textbf{2.00}&           -13.9947 &\textbf{3.00}& -20.4942 &\textbf{4.00}\\
$n=3$ & \phantom{0}-3.2754 &\textbf{1.00}& \phantom{0}-9.8614 &\textbf{2.00}&           -16.9942 &\textbf{3.00}& -24.4938 &\textbf{4.00}\\
\bottomrule
\end{tabular}
\end{table}

\subsection{Hex-quad mixed-dimensional problem}\label{SS:example_codim1}

\begin{figure}
\centering
\begin{tikzpicture}
\begin{axis}[xmin= -0.1, xmax= 3.5,
             ymin= -0.1, ymax= 1.5,
             zmin= -0.25, zmax= 1.5,
             xtick={0,0.5,1.5,2.5,3},
             xticklabels={0,0.5,0.5,0.5,1},
             ytick={1},
             yticklabels={},
             ztick={1},
             zticklabels={},
             axis equal image,
             view={-30}{15},
             width=300pt,
             height=300pt,
             axis line style={draw=none},
             axis y line=middle,
             axis z line=middle,
             tick style={draw=none},
            ]
\draw[black,dashed] (0.5,1,0) -- (2.5,1,0);
\draw[black,dashed] (0.5,1,1) -- (2.5,1,1);
\addplot3[patch,
          patch type=rectangle,
          point meta=explicit,
          shader=faceted,
          faceted color=gray,
          opacity=1.,
          fill=viridispurple,
         ] table[x=x, y=y, z=z, meta=c] {mesh_right.dat};
\addplot3[patch,
          patch type=rectangle,
          point meta=explicit,
          shader=faceted,
          faceted color=gray,
          opacity=1.,
          fill=viridispurple1,
         ] table[x=x, y=y, z=z, meta=c] {mesh_right_middle.dat};
\addplot3[patch,
          patch type=rectangle,
          point meta=explicit,
          shader=faceted,
          faceted color=black,
          opacity=1.,
          fill=gray,
         ] table[x=x, y=y, z=z, meta=c] {mesh_middle.dat};
\addplot3[patch,
          patch type=rectangle,
          point meta=explicit,
          shader=faceted,
          faceted color=gray,
          opacity=1.,
          fill=viridisyellow1,
         ] table[x=x, y=y, z=z, meta=c] {mesh_left_middle.dat};
\addplot3[patch,
          patch type=rectangle,
          point meta=explicit,
          shader=faceted,
          faceted color=gray,
          opacity=1.,
          fill=viridisyellow,
         ] table[x=x, y=y, z=z, meta=c] {mesh_left.dat};
\draw[black,dashed] (0.5,0,0) -- (2.5,0,0);
\draw[black,dashed] (0.5,0,1) -- (2.5,0,1);
\draw[-latex,draw=black,shorten <= 2pt,shorten >= 2pt] (3,0,0) -- (3.5,0,0);\node at(axis cs:3.5,0,0) [anchor=north]{$x$};
\draw[-latex,draw=black,shorten <= 2pt,shorten >= 2pt] (0,1,0) -- (0,1.5,0);\node at(axis cs:0,1.5,0) [anchor=north]{$y$};
\draw[-latex,draw=black,shorten <= 2pt,shorten >= 2pt] (0,0,1) -- (0,0,1.5);\node at(axis cs:0,0,1.5) [anchor=south]{$z$};
\node at(axis cs:0.575,0,1) [anchor=south]{$\Omega^l$};
\node at(axis cs:1.575,0,1) [anchor=south]{$\Omega^i$};
\node at(axis cs:3.075,0,1) [anchor=south]{$\Omega^r$};
\end{axis}
\end{tikzpicture}
\caption{
Hex-quad mixed cell-type problem setup.
}
\label{Fi:example_codim1}
\end{figure}
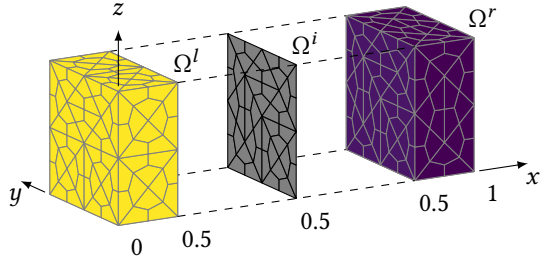

\begin{table}
\centering
\caption{Hex-quad mixed cell-type problem: parameters.\label{Ta:example_codim1_parameters}}
\begin{tabular}{ clc } 
 \toprule
Parameter & Description & Value \\ 
\midrule
$C$ & Interior penalty constant &$100.$ \\
$n$ & Mesh refinement level & $\{0,1,2,3\}$ \\
$h$ & Mesh size & $0.10/2^n$ \\
$p$ & Degree of polynomial & $\{1,2,3,4\}$ \\
\bottomrule
\end{tabular}
\end{table}

\begin{table}
\centering
\caption{Hex-quad mixed cell-type problem: solutions, test functions, and function spaces.\label{Ta:example_codim1_functions}}
\begin{tabular}{ ccccl } 
 \toprule
Solution & Test function & family & degree & Description \\ 
\midrule
$u^l\in V^l$ & $\delta u^l\in V^l_0$ & Q & $p$ & Field in $\Omega^l$\\
$u^i\in V^i$ & $\delta u^i\in V^i$   & Q & $p$ & Field in $\Omega^i$\\
$u^r\in V^r$ & $\delta u^r\in V^r_0$ & Q & $p$ & Field in $\Omega^r$\\
\bottomrule
\end{tabular}
\end{table}

In addition to the quad-triangle mixed cell-type problem
presented in \Cref{SS:example_mixed},
we solved a hex-quad mixed cell-type problem, and
performed a convergence study.
Note that our UFL abstraction for multi-domain problems
using \code{MeshSequence} \cref{SS:ufl_multi}
works seamlessly in the existence of codim-1 submeshes,
making the most of the existing implementations in UFL
for solving PDEs on embedded surfaces \citep{Rognes2013}.

\Cref{Fi:example_codim1} shows
two disjoint three-dimensional domains, $\Omega^l$ and $\Omega^r$,
that compose the unit cube, and
one two-dimensional domain, $\Omega^i$, defined at the interface of the two.

Similarly to \cref{SS:example_mixed},
we solve the Poisson problem in $\Omega^l$ and $\Omega^r$
with Dirichlet boundary conditions on
$\partial\Omega^l\setminus\Omega^i$ and
$\partial\Omega^r\setminus\Omega^i$,
but
we weakly enforce continuity of the solution and flux
by introducing an auxiliary variable and solving some equation on $\Omega^i$.
The resulting system is equivalent to one
that one would obtain with the classical interior penalty method
once we eliminate the auxiliary variable on $\Omega^i$.
Specifically, we solve the following:

$  $\\
Find $(u^l,u^i,u^r)\in V^l\times V^i\times V^r$ such that:
\begin{align}
&\int_{\Omega^l}
\nabla u^l\!\cdot\!\nabla\delta u^l\dif V+
\int_{\Omega^r}
\nabla u^r\!\cdot\!\nabla\delta u^r\dif V\notag\\
&-\int_{\Omega^i}u^i\!\cdot\!(\delta u^l-\delta u^r)\dif A
-\int_{\Omega^i}\jump{u}\!\cdot\!\avg{\nabla{\delta u}}\dif A
+\frac{C}{h}\int_{\Omega^i}\jump{u}\jump{\delta u}\dif A\notag\\
&-\int_{\Omega^i}((\nabla u^l\!\cdot\!\bs{n}^l-\nabla u^r\!\cdot\!\bs{n}^r)/2-u^i)\!\cdot\!\delta u^i\dif A\notag\\
&=
\int_{\Omega^l}f\delta u^l\dif V +
\int_{\Omega^r}f\delta u^r\dif V
\quad
\forall(\delta u^l,\delta u^i,\delta u^r)\in V^l_0\times V^i\times V^r_0,\\
u^l&=u_e\quad\text{on }\partial\Omega^l\setminus\Omega^i,\\
u^r&=u_e\quad\text{on }\partial\Omega^r\setminus\Omega^i,
\label{E:example_codim1}
\end{align}
where $\avg{\cdot}$ and $\jump{\cdot}$ operators are defined as:
\begin{align}
\avg{\nabla u}&=(\nabla u^l+\nabla u^r)/2,\\
\jump{u}&=u^l\bs{n}^l+u^r\bs{n}^r,
\end{align}
and:
\begin{align}
u_e=\cos(2\pi x)\cos(2\pi y)\cos(2\pi z),
\end{align}
and:
\begin{align}
f=12\pi^2u_e.
\end{align}
\Cref{Ta:example_codim1_parameters} summarises the parameters and
\Cref{Ta:example_codim1_functions} summarises solutions, test functions, and function spaces on which they are defined.
This is a manufactured problem whose solution is known to be $u_e$.

\cref{Fi:example_codim1} shows
hexahedral meshes of $\Omega^l$ and $\Omega^r$ and
a quadrilateral mesh of $\Omega^i$.
We solved the Poisson problem
for each refinement level $n\in\{0,1,2,3\}$
for each polynomial degree $p\in\{1,2,3,4\}$,
and performed a convergence study by computing the $L^2$- and $H^1$-norm errors of the solutions defined, respectively, as:
\begin{subequations}
\begin{align}
&(\|u^l-u_e\|_{L^2(\Omega^l)}^2+\|u^r-u_e\|_{L^2(\Omega^r)}^2)^{1/2},\\
&(\|u^l-u_e\|_{H^1(\Omega^l)}^2+\|u^r-u_e\|_{H^1(\Omega^r)}^2)^{1/2}.
\end{align}
\end{subequations}
Refinement level $n$ corresponds to the mesh
obtained by uniformly refining the mesh shown in
\cref{Fi:example_codim1} $n$ times.
We used PCFIELDSPLIT preconditioner in conjunction with conjugate gradient method with Jacobi preconditioner in PETSc KSP linear solver
to first eliminate system for $u^i$ and then solve the resulting symmetric system.
\Cref{Ta:example_codim1_L2error,Ta:example_codim1_H1error}
summarise the $\log_2$ of $L^2$- and $H^1$-norm errors, 
where we observe the expected convergence behaviour.

\begin{table}
\centering
\caption{Hex-quad mixed cell-type problem: $\log_2$ of $L^2$-norm errors for polynomial degrees $p\in\{1,2,3,4\}$ and refinement levels $n\in\{0,1,2,3\}$.\label{Ta:example_codim1_L2error}}
\begin{tabular}{ crrrrrrrr } 
 \toprule
      & $p=1$\phantom{0} &\textbf{rate}& $p=2$\phantom{0} &\textbf{rate}& $p=3$\phantom{0} &\textbf{rate}& $p=4$\phantom{0} &\textbf{rate}\\ 
\midrule
$n=0$ & \phantom{0}-3.8413 &             & \phantom{0}-7.6170 &             & -11.4946 &             & -15.5054 &\\
$n=1$ & \phantom{0}-5.7008 &\textbf{1.86}&           -10.8262 &\textbf{3.21}& -15.4673 &\textbf{3.97}& -20.6226 &\textbf{5.12}\\
$n=2$ & \phantom{0}-7.6514 &\textbf{1.95}&           -13.8869 &\textbf{3.06}& -19.4888 &\textbf{4.02}& -25.6459 &\textbf{5.02}\\
$n=3$ & \phantom{0}-9.6347 &\textbf{1.98}&           -16.9051 &\textbf{3.02}& -23.5066 &\textbf{4.02}& -30.6549 &\textbf{5.01}\\
\bottomrule
\end{tabular}
\end{table}
\begin{table}
  \centering
\caption{Hex-quad mixed cell-type problem: $\log_2$ of $H^1$-norm errors for polynomial degrees $p\in\{1,2,3,4\}$ and refinement levels $n\in\{0,1,2,3\}$.\label{Ta:example_codim1_H1error}}
\begin{tabular}{ crrrrrrrr } 
 \toprule
      & $p=1$\phantom{0} &\textbf{rate}& $p=2$\phantom{0} &\textbf{rate}& $p=3$\phantom{0} &\textbf{rate}& $p=4$\phantom{0} &\textbf{rate}\\ 
\midrule
$n=0$ & \phantom{0-}0.4687 &             & \phantom{0}-2.1871 &             & \phantom{0}-5.4461 &             & \phantom{0}-9.1030 &\\
$n=1$ & \phantom{0}-0.4645 &\textbf{0.93}& \phantom{0}-4.2736 &\textbf{2.09}& \phantom{0}-8.3493 &\textbf{2.90}&           -13.1866 &\textbf{4.08}\\
$n=2$ & \phantom{0}-1.4284 &\textbf{0.96}& \phantom{0}-6.2878 &\textbf{2.01}&           -11.3185 &\textbf{2.97}&           -17.1989 &\textbf{4.01}\\
$n=3$ & \phantom{0}-2.4156 &\textbf{0.99}& \phantom{0}-8.2876 &\textbf{2.00}&           -14.3080 &\textbf{2.99}&           -21.1991 &\textbf{4.00}\\
\bottomrule
\end{tabular}
\end{table}

\subsection{Fluid-structure interaction problem}\label{SS:example_fsi}
We solved the fluid-structure interaction (FSI) benchmark problem proposed in \citet{TurekHron2006, TurekHron2010},
and compared the results with the reference.
The code to reproduce the results in this section is found at \cite{firedrake_zenodo}. 

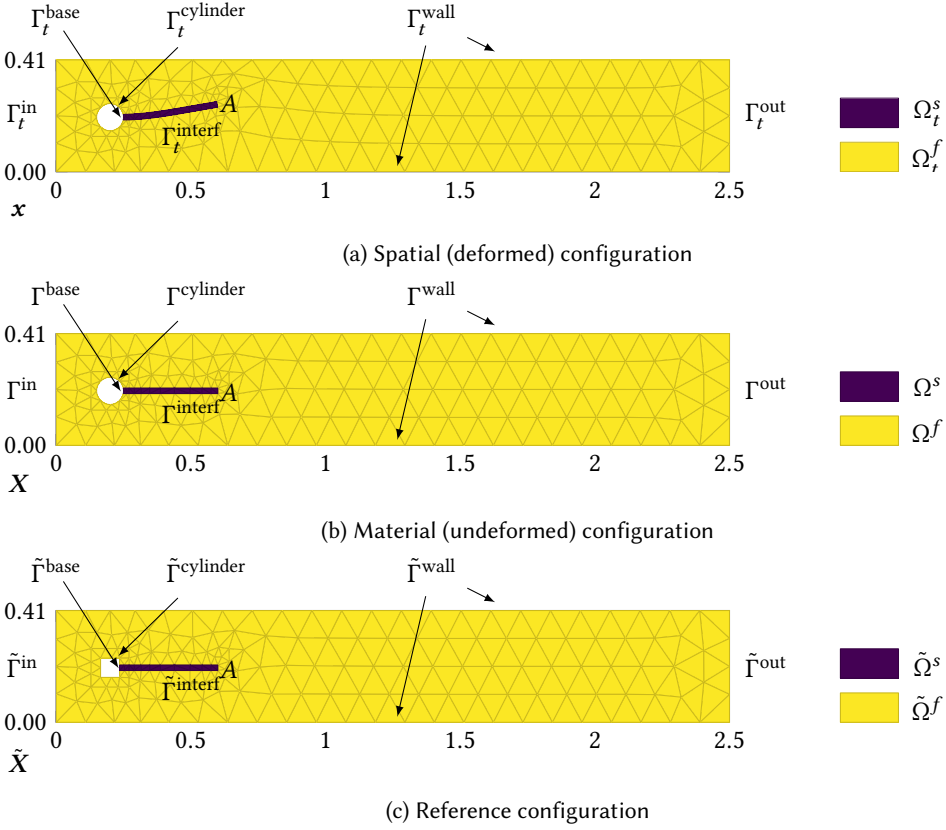
\begin{figure}[ht]
\centering
\begin{subfigure}[b]{1.0\textwidth}
\centering
\begin{tikzpicture}
\begin{axis}[
             xmin= 0.00, xmax= 2.50,
             ymin= 0.00, ymax= 0.41,
             point meta min= 0., point meta max= 1.0,
             axis equal image,
             colormap/viridis, 
             width=300pt,
             height=200pt,
             ytick={0.0, 0.41},
             yticklabels={0.00, 0.41},
            ]
\addplot[patch,
          patch type=triangle quadr,
          point meta=explicit,
          shader=faceted interp, 
          opacity=1.,
         ] table[x=x, y=y, meta=mesh_f] {mesh_f_spatial.dat};
\addplot[patch,
          patch type=triangle quadr,
          point meta=explicit,
          shader=faceted interp, 
          opacity=1.,
         ] table[x=x, y=y, meta=mesh_s] {mesh_s_spatial.dat};
\end{axis}
\node[] (anchor=west) at (-0.45, 0.75) {$\Gamma^{\text{in}}_t$};
\node[] (anchor=east) at (9.45, 0.75) {$\Gamma^{\text{out}}_t$};
\node[] (interf) at (1.8, 0.45) {$\Gamma^{\text{interf}}_t$};
\node[] () at (2.3, 0.90) {$A$};
\node[] (wall_start) at (5., 2.) {};
\node[] (wall_end0) at (6., 1.5) {};
\node[] (wall_end1) at (4.5, -0.1) {};
\node[] (anchor=west) at ($(wall_start)$) {$\Gamma^{\text{wall}}
_t$};
\path[draw=black,-latex,shorten <= 15pt,shorten >= 5pt] ($(wall_start)$)--($(wall_end0)$);
\path[draw=black,-latex,shorten <= 10pt,shorten >= 5pt] ($(wall_start)$)--($(wall_end1)$);
\node[] (cylinder_start) at (2., 2.) {};
\node[] (cylinder_end) at (0.70, 0.75) {};
\node[] (anchor=west) at ($(cylinder_start)$) {$\Gamma^{\text{cylinder}}
_t$};
\path[draw=black,-latex,shorten <= 15pt,shorten >= 5pt] ($(cylinder_start)$)--($(cylinder_end)$);
\node[] (base_start) at (0., 2.) {};
\node[] (base_end) at (0.90, 0.65) {};
\node[] (anchor=east) at ($(base_start)$) {$\Gamma^{\text{base}}_t$};
\path[draw=black,-latex,shorten <= 5pt,shorten >= 2pt] ($(base_start)$)--($(base_end)$);
\node[] () at (-0.5, -0.5) {$\bs{x}$};
\begin{axis}[
             xmin= 0.00, xmax= 4.50,
             ymin= 0.00, ymax= 0.41,
             point meta min= 0., point meta max= 1.0,
             axis equal image,
             colormap/viridis,
             width=540pt,
             height=200pt,
             xtick={},
             xticklabels={},
             ytick={},
             yticklabels={},
             axis line style={draw=none},
             tick style={draw=none},
            ]
\addplot[patch,
         patch type=rectangle,
         point meta=explicit,
         shader=faceted interp,
         opacity=1.,
        ] table[x=x, y=y, meta=c, row sep=crcr] {
x y c\\
2.7 0.0 1.\\
2.9 0.0 1.\\
2.9 0.1 1.\\
2.7 0.1 1.\\
};
\addplot[patch,
         patch type=rectangle,
         point meta=explicit,
         shader=faceted interp,
         opacity=1.,
        ] table[x=x, y=y, meta=c, row sep=crcr] {
x y c\\
2.7 0.15 0.\\
2.9 0.15 0.\\
2.9 0.25 0.\\
2.7 0.25 0.\\
};
\node[] (anchor=south east) at (3.0, 0.20) {$\Omega^s_t$};
\node[] (anchor=south east) at (3.0, 0.05) {$\Omega^f_t$};
\end{axis}
\end{tikzpicture}
\caption{Spatial (deformed) configuration}
\label{Fi:example_fsi_setting_spatial}
\end{subfigure}

\begin{subfigure}[b]{1.0\textwidth}
\centering
\begin{tikzpicture}
\begin{axis}[
             xmin= 0.00, xmax= 2.50,
             ymin= 0.00, ymax= 0.41,
             point meta min= 0., point meta max= 1.0,
             axis equal image,
             colormap/viridis, 
             width=300pt,
             height=200pt,
             ytick={0.0, 0.41},
             yticklabels={0.00, 0.41},
            ]
\addplot[patch,
          patch type=triangle quadr,
          point meta=explicit,
          shader=faceted interp, 
          opacity=1.,
         ] table[x=x, y=y, meta=mesh_f] {mesh_f_material.dat};
\addplot[patch,
          patch type=triangle quadr,
          point meta=explicit,
          shader=faceted interp, 
          opacity=1.,
         ] table[x=x, y=y, meta=mesh_s] {mesh_s_material.dat};
\end{axis}
\node[] (anchor=west) at (-0.45, 0.75) {$\Gamma^{\text{in}}$};
\node[] (anchor=east) at (9.45, 0.75) {$\Gamma^{\text{out}}$};
\node[] (interf) at (1.8, 0.45) {$\Gamma^{\text{interf}}$};
\node[] () at (2.3, 0.70) {$A$};
\node[] (wall_start) at (5., 2.) {};
\node[] (wall_end0) at (6., 1.5) {};
\node[] (wall_end1) at (4.5, -0.1) {};
\node[] (anchor=west) at ($(wall_start)$) {$\Gamma^{\text{wall}}$};
\path[draw=black,-latex,shorten <= 15pt,shorten >= 5pt] ($(wall_start)$)--($(wall_end0)$);
\path[draw=black,-latex,shorten <= 10pt,shorten >= 5pt] ($(wall_start)$)--($(wall_end1)$);
\node[] (cylinder_start) at (2., 2.) {};
\node[] (cylinder_end) at (0.70, 0.75) {};
\node[] (anchor=west) at ($(cylinder_start)$) {$\Gamma^{\text{cylinder}}
$};
\path[draw=black,-latex,shorten <= 15pt,shorten >= 5pt] ($(cylinder_start)$)--($(cylinder_end)$);
\node[] (base_start) at (0., 2.) {};
\node[] (base_end) at (0.90, 0.65) {};
\node[] (anchor=east) at ($(base_start)$) {$\Gamma^{\text{base}}$};
\path[draw=black,-latex,shorten <= 5pt,shorten >= 2pt] ($(base_start)$)--($(base_end)$);
\node[] () at (-0.5, -0.5) {$\bs{X}$};
\begin{axis}[
             xmin= 0.00, xmax= 4.50,
             ymin= 0.00, ymax= 0.41,
             point meta min= 0., point meta max= 1.0,
             axis equal image,
             colormap/viridis,
             width=540pt,
             height=200pt,
             xtick={},
             xticklabels={},
             ytick={},
             yticklabels={},
             axis line style={draw=none},
             tick style={draw=none},
            ]
\addplot[patch,
         patch type=rectangle,
         point meta=explicit,
         shader=faceted interp,
         opacity=1.,
        ] table[x=x, y=y, meta=c, row sep=crcr] {
x y c\\
2.7 0.0 1.\\
2.9 0.0 1.\\
2.9 0.1 1.\\
2.7 0.1 1.\\
};
\addplot[patch,
         patch type=rectangle,
         point meta=explicit,
         shader=faceted interp,
         opacity=1.,
        ] table[x=x, y=y, meta=c, row sep=crcr] {
x y c\\
2.7 0.15 0.\\
2.9 0.15 0.\\
2.9 0.25 0.\\
2.7 0.25 0.\\
};
\node[] (anchor=south east) at (3.0, 0.20) {$\Omega^s$};
\node[] (anchor=south east) at (3.0, 0.05) {$\Omega^f$};
\end{axis}
\end{tikzpicture}
\caption{Material (undeformed) configuration}
\label{Fi:example_fsi_setting_material}
\end{subfigure}

\begin{subfigure}[b]{1.0\textwidth}
\centering
\begin{tikzpicture}
\begin{axis}[
             xmin= 0.00, xmax= 2.50,
             ymin= 0.00, ymax= 0.41,
             point meta min= 0., point meta max= 1.0,
             axis equal image,
             colormap/viridis,
             width=300pt,
             height=200pt,
             ytick={0.0, 0.41},
             yticklabels={0.00, 0.41},
            ]
\addplot[patch,
          patch type=triangle quadr,
          point meta=explicit,
          shader=faceted interp, 
          opacity=1.,
         ] table[x=x, y=y, meta=mesh_f] {mesh_f_reference.dat};
\addplot[patch,
          patch type=triangle quadr,
          point meta=explicit,
          shader=faceted interp, 
          opacity=1.,
         ] table[x=x, y=y, meta=mesh_s] {mesh_s_reference.dat};
\end{axis}
\node[] (anchor=west) at (-0.45, 0.75) {$\tilde{\Gamma}^{\text{in}}$};
\node[] (anchor=east) at (9.45, 0.75) {$\tilde{\Gamma}^{\text{out}}$};
\node[] (interf) at (1.8, 0.45) {$\tilde{\Gamma}^{\text{interf}}$};
\node[] () at (2.3, 0.70) {$A$};
\node[] (wall_start) at (5., 2.) {};
\node[] (wall_end0) at (6., 1.5) {};
\node[] (wall_end1) at (4.5, -0.1) {};
\node[] (anchor=west) at ($(wall_start)$) {$\tilde{\Gamma}^{\text{wall}}$};
\path[draw=black,-latex,shorten <= 15pt,shorten >= 5pt] ($(wall_start)$)--($(wall_end0)$);
\path[draw=black,-latex,shorten <= 10pt,shorten >= 5pt] ($(wall_start)$)--($(wall_end1)$);
\node[] (cylinder_start) at (2., 2.) {};
\node[] (cylinder_end) at (0.70, 0.75) {};
\node[] (anchor=west) at ($(cylinder_start)$) {$\tilde{\Gamma}^{\text{cylinder}}
$};
\path[draw=black,-latex,shorten <= 15pt,shorten >= 5pt] ($(cylinder_start)$)--($(cylinder_end)$);
\node[] (base_start) at (0., 2.) {};
\node[] (base_end) at (0.87, 0.65) {};
\node[] (anchor=east) at ($(base_start)$) {$\tilde{\Gamma}^{\text{base}}$};
\path[draw=black,-latex,shorten <= 5pt,shorten >= 2pt] ($(base_start)$)--($(base_end)$);
\node[] () at (-0.5, -0.5) {$\tilde{\bs{X}}$};
\begin{axis}[
             xmin= 0.00, xmax= 4.50,
             ymin= 0.00, ymax= 0.41,
             point meta min= 0., point meta max= 1.0,
             axis equal image,
             colormap/viridis,
             width=540pt,
             height=200pt,
             xtick={},
             xticklabels={},
             ytick={},
             yticklabels={},
             axis line style={draw=none},
             tick style={draw=none},
            ]
\addplot[patch,
         patch type=rectangle,
         point meta=explicit,
         shader=faceted interp,
         opacity=1.,
        ] table[x=x, y=y, meta=c, row sep=crcr] {
x y c\\
2.7 0.0 1.\\
2.9 0.0 1.\\
2.9 0.1 1.\\
2.7 0.1 1.\\
};
\addplot[patch,
         patch type=rectangle,
         point meta=explicit,
         shader=faceted interp,
         opacity=1.,
        ] table[x=x, y=y, meta=c, row sep=crcr] {
x y c\\
2.7 0.15 0.\\
2.9 0.15 0.\\
2.9 0.25 0.\\
2.7 0.25 0.\\
};
\node[] (anchor=south east) at (3.0, 0.20) {$\tilde{\Omega}^s$};
\node[] (anchor=south east) at (3.0, 0.05) {$\tilde{\Omega}^f$};
\end{axis}
\end{tikzpicture}
\caption{Reference configuration}
\label{Fi:example_fsi_setting_reference}
\end{subfigure}
\caption{ FSI benchmark problem setup. The meshes are at vastly reduced
resolution for legibility.}
\label{Fi:example_fsi_setting}
\end{figure}

\Cref{Fi:example_fsi_setting} shows the domain of interest in three different configurations:
the time-dependent spatial (undeformed) configuration
(\cref{Fi:example_fsi_setting_spatial}),
the time-independent material (deformed) configuration
(\cref{Fi:example_fsi_setting_material}), and
the reference configuration
(\cref{Fi:example_fsi_setting_reference}),
and introduces notation.

We solve an incompressible Navier-Stokes equation for a Newtonian fluid
in the fluid domain
for fluid velocity $\bs{v}^f$ and pressure $p^f$ and
a finite-strain nonlinear elasticity problem
with a St.\ Venant-Kirchhoff material model
in the structure domain
for structure velocity $\bs{v}^s$ and displacement $\bs{u}^s$,
written in first-order-in-time form.
Equations for fluids are normally written in the spatial configuration and
those for structures are normally written in the material configuration.
As the structure moves in time, the domain on which
the fluid problem is defined changes.
We use the arbitrary Lagrangian-Eulerian (ALE) method \citep{Hirt1974, Belytschko1978, Hughes1981},
to recast the fluid problem in the spatial configuration as one in the material configuration
to solve the coupled system in a time-independent frame.
ALE methods require one to solve an artificial boundary value problem to represent the fluid mesh motion $\bs{u}^f$.
In this example,
we solve a fictitious vector biharmonic equation for $\bs{u}^f$
\citep{Helenbrook2003},
using $H^2$-conforming spaces for $\bs{u}^s$ and $\bs{v}^s$ as well as for $\bs{u}^f$.
For the reason clarified in Sec.~\cref{SSS:example_fsi_disc_space},
we further recast the coupled system written in the material configuration as one in the \emph{reference} configuration
in which cylindrical boundary is \emph{flattened}
to align with the coordinate axes;
see \cref{Fi:example_fsi_setting_reference}.
We solve the numerical problem in the reference configuration.

We denote the coordinates in the spatial, material, and reference configurations by 
$\bs{x}$, $\bs{X}$, and $\tilde{\bs{X}}$.
We have:
\begin{align}
\bs{x}&=
\begin{cases}
	\bs{X}+\bs{u}^f & \text{ in fluid},\\
    \bs{X}+\bs{u}^s & \text{ in structure},
\end{cases}\\
\bs{X}&=\tilde{\bs{X}}+\bs{u}^m,
\end{align}
where $\bs{u}^f$ and $\bs{u}^s$ are enforced to be continuous across the fluid-structure interface by boundary conditions and
$\bs{u}^m$ is the known \emph{displacement} field
from the reference configuration to the material configuration
that is continuous across the fluid-structure interface.

In the spatial, material, and reference configurations,
we denote
by $\dif v$, $\dif V$, and $\dif \tilde{V}$ the volume measures and
by $\dif a$, $\dif A$, and $\dif \tilde{A}$ the surface measures,
Furthermore, we denote
by $\bs{n}^f$, $\bs{N}^f$, and $\tilde{\bs{N}}^f$ the outward normal to the fluid domain and
by $\bs{n}^s$, $\bs{N}^s$, and $\tilde{\bs{N}}^s$ the outward normal to the structure domain.

\Cref{Ta:example_fsi_parameters} summarises the parameters and
\cref{Ta:example_fsi_functions} summarises solutions, test functions, and function spaces in the reference configuration on which they are defined.
Equations and boundary conditions are summarised in the following.

\begin{table}
\centering
\caption{FSI problem: parameters.\label{Ta:example_fsi_parameters}}
\begin{tabular}{ clc } 
 \toprule
Parameter & Description & Value \\ 
\midrule
$L$ & Length of the domain &$2.5$ \\
$H$ & Height of the domain & $0.41$ \\ 
$C$ & Center of the cylinder & $(0.20,0.20)$\\ 
$r$ & Radius of the cylinder & $0.050$\\
$A$ & Tip of the structure (reference point) & $(0.60,0.20)$\\
$h$ & Thickness of the structure & $0.020$\\
$T$ & Duration of the simulation & $20.$\\
$\rho^f$ & Density of the fluid & $1.0\cdot 10^{3}$\\ 
$\nu^f$ & Poisson's ratio of the fluid & $1.0\cdot 10^{\minus 3}$\\
$\rho^s$ & Density of the structure & $1.0\cdot 10^{3}$\\ 
$\nu^s$ & Poisson's ratio of the structure & $0.40$\\ 
$\mu^s$ & Lam\'{e} parameter & $2.0\cdot 10^{6}$\\ 
$\lambda^s$ & Lam\'{e} parameter & $\mu^s\cdot2\nu^s/(1\minus 2\nu^s)$\\ 
$\bar{\bs{v}}^f_{\text{inflow}}$ & Fluid inflow constant & $2.0$\\ 
\bottomrule
\end{tabular}
\end{table}

\begin{table}
\centering
\caption{FSI problem: solutions, test functions, and function spaces in the reference configuration.\label{Ta:example_fsi_functions}}
\begin{tabular}{ ccccl } 
 \toprule
Solution & Test function & family & degree & Description \\ 
\midrule
$\bs{v}^f\in V^f$ & $\delta\bs{v}^f\in V^f_0$ & P & 5 & Fluid velocity\\
$\bs{u}^f\in U^f$ & $\delta\bs{u}^f\in U^f_0$ & Argyris & 5 & Fluid mesh motion\\
$p^f\in Q^f$ & $\delta p^f\in Q^f$ & P & 3 & Fluid pressure\\
$\bs{v}^s\in V^s$ & $\delta\bs{v}^s\in V^s_0$ & Argyris & 5 & Structure velocity\\
$\bs{u}^s\in U^s$ & $\delta\bs{u}^s\in U^s_0$ & Argyris & 5 & Structure displacement\\
\bottomrule
\end{tabular}
\end{table}

\subsubsection{Equations for fluid velocity $\bs{v}^f$}\label{SSS:example_fsi_vf}
Equations for the fluid velocity $\bs{v}^f\in V^f$ are written in the spatial configuration as:
\begin{align}
\int_{\Omega^f_t}\left(
\rho^f\dot{\bs{v}^f}\!\cdot\!\delta\bs{v}^f
+\rho^f\grad\bs{v}^f(\bs{v}^f-\dot{\bs{u}^f})\!\cdot\!\delta\bs{v}^f
+\bs{\sigma}^f\!:\!\grad\delta\bs{v}^f
\right)\dif v = 0\quad\forall\delta\bs{v}^f\in V^f_0,
\label{E:example_fsi_dvf}
\end{align}
where:
\begin{align}
\bs{\sigma}^f&:=-p^f\bs{I}+\rho^f\nu^f(\grad\bs{v}^f+(\grad\bs{v}^f)^{\text{T}}).
\end{align}
Dirichlet boundary conditions are given as:
\begin{align}
\bs{v}^f &= \bs{v}^f_{\text{inflow}}\quad\on\Gamma^{\text{in}}_t,\\
\bs{v}^f &= \bs{0}\quad\on\Gamma^{\text{wall}}_t\cup\Gamma^{\text{cylinder}}_t,\\
\bs{v}^f &= \bs{v}^s\quad\on\Gamma^{\text{interf}}_t,
\end{align}
where:
\begin{align}
\bs{v}^f_{\text{inflow}}=
1.5\cdot\bar{\bs{v}}^f_{\text{inflow}}\cdot\frac{y(H-y)}{(H/2)^2}
\begin{cases}
	\frac{1-\cos(\pi/2\cdot t)}{2}, & t<2,\\
    1, & \text{otherwise.}
\end{cases}
\end{align}

\subsubsection{Equations for fluid mesh motion $\bs{u}^f$}\label{SSS:example_fsi_uf}
Equations for the fluid mesh motion $\bs{u}^f\in U^f$ are written in the material configuration as:
\begin{align}
\int_{\Omega^f}\Grad^2\bs{u}^f\!\cdot\!\Grad^2\delta\bs{u}^f\dif V &= 0
\quad\forall\delta\bs{u}^f\in U^f_0,
\label{E:example_fsi_duf}
\end{align}
where homogeneous high-order Neumann boundary conditions were applied.
Dirichlet boundary conditions are given as:
\begin{align}
\bs{u}^f&=\bs{0}\quad\on\Gamma^{\text{in}}\cup\Gamma^{\text{out}}\cup\Gamma^{\text{wall}}\cup\Gamma^{\text{cylinder}},\\
\bs{u}^f&=\bs{u}^s\quad\on\Gamma^{\text{interf}}.
\end{align}

\subsubsection{Equation for fluid pressure $p^f$}\label{SSS:example_fsi_pf}
Equation for the fluid pressure $p^f\in Q^f$ is written in the spatial configuration as:
\begin{align}
\int_{\Omega^f_t}\div\bs{v}^f\delta p^f\dif v=0\quad\forall\delta p^f\in Q^f.
\label{E:example_fsi_dpf}
\end{align}

\subsubsection{Equations for structural velocity $\bs{v}^s$}\label{SSS:example_fsi_vs}
Equations for the structural velocity $\bs{v}^s\in V^s$
are written in the material configuration as:
\begin{align}
\int_{\Omega^s}\left(
(\rho^sJ^s)\dot{\bs{v}^s}\!\cdot\!\delta\bs{v}^s
+\bs{P}\!:\!\Grad\delta\bs{v}^s
\right)\dif V
&-\int_{\Gamma^{\text{interf}}}
\hspace{-5pt}\bs{t}^s\!\cdot\!\delta\bs{v}^s\dif A=0\quad\forall\delta\bs{v}^s\in V^s,
\label{E:example_fsi_dvs}
\end{align}
where:
\begin{align}
\bs{t}^s:&=J^s\bs{\sigma}^f(\bs{F}^s)^{-\text{T}}\bs{N}^s,\\
\bs{P}&:=J^s\bs{\sigma}^s(\bs{F}^s)^{-\text{T}},\\
\bs{\sigma}^s&:=\frac{1}{J^s}\bs{F}^s(\lambda^s(\trace\bs{E}^s)\bs{I}+2\mu^s\bs{E}^s)(\bs{F}^{s})^{\text{T}},\\
\bs{E}^s&:=\frac{1}{2}((\bs{F}^{s})^{\text{T}}\bs{F}^s-\bs{I}),\\
\bs{F}^s&:=\Grad(\bs{X}+\bs{u}^s),\\
J^s&:=\determinant\bs{F}^s.
\end{align}
Dirichlet boundary conditions are given as:
\begin{align}
\bs{v}^s&=\bs{0}\quad\on\Gamma^{\text{base}}_t.
\end{align}

\subsubsection{Equations for structural displacement $\bs{u}^s$}\label{SSS:example_fsi_us}
Equations for the structural displacement $\bs{u}^s\in U^s$ are given as:
\begin{align}
\int_{\Omega^s_t}\left(
\dot{\bs{u}^s}-\bs{v}^s
\right)
\!\cdot\!\delta\bs{u}^s
\dif v = 0.
\label{E:example_fsi_dus}
\end{align}
Dirichlet boundary conditions are given as:
\begin{align}
\bs{u}^s=\bs{0}\quad\on\Gamma^{\text{base}}.
\end{align}

\subsubsection{Changes of coordinates}\label{SSS:example_fsi_coords}
Equations for fluid velocity and pressure
written in the spatial configuration are
first rewritten in the material configuration using the following transformation rules:
\begin{align}
\dif v&=\det(\Grad\bs{x})\dif V,\\
\grad(\cdot)&=\Grad(\cdot)\cdot(\Grad\bs{x})^{-1}.
\end{align}
Equations written in the material configuration
are then rewritten in the reference configuration
using the following transformation rules:
\begin{align}
\dif V&=\det(\rgrad\bs{X})\dif\tilde{V},\\
\Grad(\cdot)&=\rgrad(\cdot)\cdot(\rgrad\bs{X})^{-1}.
\end{align}

\subsubsection{Monolithic residual and Jacobian}\label{SSS:example_fsi_monolithic}
Let:
\begin{align}
V=V^f\times U^s\times Q^f\times V^s\times U^s,\\
V_0=V^f_0\times U^s_0\times Q^f\times V^s_0\times U^s_0,
\end{align}
and define $u\in V$ and $\delta u\in V_0$ monolithically as:
\begin{align}
u:=(\bs{v}^f,\bs{u}^f,p^f,\bs{v}^s,\bs{u}^s),\\
\delta u:=(\delta\bs{v}^f,\delta\bs{u}^f,\delta p^f,\delta\bs{v}^s,\delta\bs{u}^s).
\end{align}
Then, one can sum
\cref{E:example_fsi_dvf},
\cref{E:example_fsi_duf},
\cref{E:example_fsi_dpf},
\cref{E:example_fsi_dvs}, and
\cref{E:example_fsi_dus}
to obtain a monolithic representation of residual, $F=F(u;\delta u)$, and then
a monolithic representation of Jacobian, $J=\delta_{u}F(u;\delta u)$.

\subsubsection{Spatial discretisation}\label{SSS:example_fsi_disc_space}
We used ngsPETSc \cite{ngspetsc} to discretise the domain in the reference configuration
to obtain the mesh shown in \cref{Fi:example_fsi_setting_reference};
the corresponding meshes in the material and the spatial configurations
are shown in \cref{Fi:example_fsi_setting_material} and \cref{Fi:example_fsi_setting_spatial}.
The mesh was generated so that the fluid-structure interface $\tilde{\Gamma}^{\text{interf}}$ would not cut through cells.
The mesh shown in \cref{Fi:example_fsi_setting_reference}
was then uniformly refined three times.
We then extracted
the mesh of the fluid domain $\tilde{\Omega}^f$ and
that of the structure domain $\Omega^s$
as submeshes.

The finite element function spaces and functions that we used in this example are summarised in \cref{Ta:example_fsi_functions}.
In both \citet{TurekHron2006} and \citet{TurekHron2010} a
discretisation based on the classical Q2/P1 finite element pair was used;
total number of DoFs are compared in \cref{Ta:example_fsi_disc}

We solved the numerical problem in the reference configuration,
not directly in the material configuration,
to apply Dirichlet boundary conditions strongly on the Argyris spaces.

\subsubsection{Temporal discretisation}\label{SSS:example_fsi_disc_time}
We denote by $(\cdot)^n$ the quantity at $t=t^n$.
We used a variant of the shifted Crank–Nicolson method,
see, for example, \citet{Richter2017},
for time-integration, and solved at each $t=t^n$:
\begin{align}
\left(\frac{1}{2}+\theta\cdot\Delta t\right)F^{n+1}+
\left(\frac{1}{2}-\theta\cdot\Delta t\right)F^{n}=0,
\label{E:example_fsi_CN}
\end{align}
with $\theta=1$, where, in both $F^{n+1}$ and $F^n$,
$\dot{\bs{u}^f}$ in \cref{E:example_fsi_dvf},
as well as the other time-derivative terms,
was approximated by the central-difference method as:
\begin{align}
\dot{\bs{u}^f}\approx\frac{\bs{u}^{f,n+1}-\bs{u}^{f,n}}{\Delta t}.
\end{align}
One can readily show that the time-integrator \cref{E:example_fsi_CN} is second-order in time.
In \citet{TurekHron2006, TurekHron2010} the standard Crank-Nicolson method was used;
timestep sizes are compared in \cref{Ta:example_fsi_disc}.

\subsection{Results and comparison with the reference}
\begin{figure}
    \centering
    \includegraphics[scale=1]{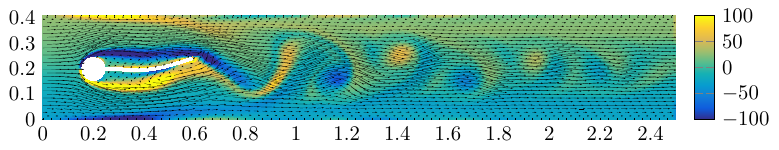}
    \caption{
Snapshot of fluid velocity $\bs{v}^f$ and vorticity $\bs{\nabla}\times\bs{v}^f$ at $t=4.11$.
}
\label{Fi:example_fsi_solution}
\end{figure}

\begin{table}
\centering
\caption{Total number of DoFs and timestep size.}
\begin{tabular}{ ccc } 
 \toprule
                      & Num. DoFs & timestep size \\ 
 \midrule
 \citet{TurekHron2006} & 304,128 & 0.00050 \\
 \citet{TurekHron2010} & 304,128 & 0.00025 \\ 
 \textbf{Our example}  & 754,260 & 0.00100 \\ 
 \bottomrule
\end{tabular}
\label{Ta:example_fsi_disc}
\end{table}
The simulation was run until $T=20$.
We used the MUMPS parallel sparse direct solver
to solve the linearised system inside Newton iterations in PETSc.
\Cref{Fi:example_fsi_solution} shows a snapshot of fluid velocity $\bs{v}$ and vorticity $\bs{\nabla}\times\bs{v}^f$ at $t=4.11$.

At each time $t$,
we computed $x$ and $y$ displacements at reference point $A$,
$u^A_x$ and $u^A_y$, and
drag and lift forces acting on
the structure-cylinder unit on
$\Gamma^{\text{interf}}_t\cup\Gamma^{\text{cylinder}}_t$,
$F_D$ and $F_L$, defined as:
\begin{align}
(F_D, F_L) =-
\int_{\Gamma^{\text{interf}}_t\cup\Gamma^{\text{cylinder}}_t}
\bs{\sigma}^f\bs{n}^f\dif a.
\end{align}
\Cref{Fi:example_fsi_solution_macro} shows
plots of
\subref{Fi:example_fsi_solution_macro_ux}
$u^A_x$,
\subref{Fi:example_fsi_solution_macro_uy}
$u^A_y$,
\subref{Fi:example_fsi_solution_macro_drag}
$F_D$, and
\subref{Fi:example_fsi_solution_macro_lift}
$F_L$
against time $t$ for $\Delta t = \{0.002, 0.001\}$.
These plots compare well with those presented in \citet{TurekHron2006}
except that one can observe minor phase leads;
given that we observe no notable difference for
$\Delta t=0.002$ and $\Delta t=0.001$,
this is presumably due to that we use higher-order elements.
\Cref{Ta:example_fsi_macro} compares mean $\pm$ amplitude values
of $u^A_x$, $u^A_y$, $F_D$, and $F_L$
with \citet{TurekHron2006,TurekHron2010}, and
shows good agreement.

\begin{figure}[ht]
\begin{subfigure}[b]{90pt}
\begin{tikzpicture}
\begin{axis}[
             xmin=19.5, xmax=20.,
             ymin=-0.006, ymax=0.,
             line width=1.0pt,
             width=110pt,
             height=110pt,
            ]
\addplot[dashed, color=red, mark=o, mark size=0.0pt] table[x=t, y=val] {time_series_ux_P5_P3_Argyris5_nref3_0.002_shift1_biharmonic.dat};
\addplot[color=red, mark=o, mark size=0.0pt] table[x=t, y=val] {time_series_ux_P5_P3_Argyris5_nref3_0.001_shift1_biharmonic.dat};
\end{axis}
\end{tikzpicture}
\caption{$u^A_x$ v.s. $t$.}
\label{Fi:example_fsi_solution_macro_ux}
\end{subfigure}
\begin{subfigure}[b]{90pt}
\begin{tikzpicture}
\begin{axis}[
             xmin=19.5, xmax=20.,
             ymin=-0.04, ymax=0.04,
             line width=1.0pt,
             width=110pt,
             height=110pt,
             legend style={font=\tiny},
            ]
\addplot[dashed, color=red, mark=o, mark size=0.0pt] table[x=t, y=val] {time_series_uy_P5_P3_Argyris5_nref3_0.002_shift1_biharmonic.dat};
\addplot[color=red, mark=o, mark size=0.0pt] table[x=t, y=val] {time_series_uy_P5_P3_Argyris5_nref3_0.001_shift1_biharmonic.dat};
\end{axis}
\end{tikzpicture}
\caption{$u^A_y$ v.s. $t$.}
\label{Fi:example_fsi_solution_macro_uy}
\end{subfigure}
\begin{subfigure}[b]{90pt}
\begin{tikzpicture}
\begin{axis}[
             xmin=19.5, xmax=20.,
             ymin=430., ymax=485.,
             line width=1.0pt,
             width=110pt,
             height=110pt,
             legend columns=-1,
             legend style={
                    at={(0.25, 1.15)},
                    column sep=1ex,
                    anchor=south west,
             },
            ]
\addplot[dashed, color=red, mark=o, mark size=0.0pt] table[x=t, y=val] {time_series_FD_P5_P3_Argyris5_nref3_0.002_shift1_biharmonic.dat};
\addplot[color=red, mark=o, mark size=0.0pt] table[x=t, y=val] {time_series_FD_P5_P3_Argyris5_nref3_0.001_shift1_biharmonic.dat};
\legend{$\Delta t=0.002$, $\Delta t=0.001$}
\end{axis}
\end{tikzpicture}
\caption{$F_D$ v.s. $t$.}
\label{Fi:example_fsi_solution_macro_drag}
\end{subfigure}
\begin{subfigure}[b]{90pt}
\begin{tikzpicture}
\begin{axis}[
             xmin=19.5, xmax=20.,
             ymin=-150., ymax=200.,
             line width=1.0pt,
             width=110pt,
             height=110pt,
            ]
\addplot[dashed, color=red, mark=o, mark size=0.0pt] table[x=t, y=val] {time_series_FL_P5_P3_Argyris5_nref3_0.002_shift1_biharmonic.dat};
\addplot[color=red, mark=o, mark size=0.0pt] table[x=t, y=val] {time_series_FL_P5_P3_Argyris5_nref3_0.001_shift1_biharmonic.dat};
\end{axis}
\end{tikzpicture}
\caption{$F_L$ v.s. $t$.}
\label{Fi:example_fsi_solution_macro_lift}
\end{subfigure}
\caption{
Plots of
\subref{Fi:example_fsi_solution_macro_ux}
$u^A_x$,
\subref{Fi:example_fsi_solution_macro_uy}
$u^A_y$,
\subref{Fi:example_fsi_solution_macro_drag}
$F_D$, and
\subref{Fi:example_fsi_solution_macro_lift}
$F_L$
against time $t$ for $\Delta t = \{0.002, 0.001\}$, where
$u^A_x$ and $u^A_y$ are $x$- and $y$-displacements at reference point A
and
$F_D$ and $F_L$ are
drag and lift forces acting on the structure-cylinder unit.
}
\label{Fi:example_fsi_solution_macro}
\end{figure}

\begin{table}
\centering
\caption{Mean $\pm$ amplitude values of
$x$- and $y$-displacements at reference point A, $u^A_x$ and $u^A_y$, and
drag and lift forces acting on the structure-cylinder unit, $F_D$ and $F_L$.}
\begin{tabular}{ ccccc }
 \toprule
                      & $u^A_x$ [$10^{-3}$] & $u^A_y$[$10^{-3}$] & $F_D$ & $F_L$\\
 \midrule
 \citet{TurekHron2006} & -2.69$\pm$ 2.53 & 1.48$\pm$ 34.38 & 457.3$\pm$ 22.66 & 2.22$\pm$ 149.78 \\
 \citet{TurekHron2010} & -2.88$\pm$ 2.72 & 1.47$\pm$ 34.99 & 460.5$\pm$ 27.74 & 2.50$\pm$ 153.91 \\
 \textbf{Our example}  & \textbf{-2.96}$\pm$\textbf{2.70} & \textbf{1.46}$\pm$\textbf{35.18} & \textbf{460.3}$\pm$\textbf{27.82} & \textbf{2.31}$\pm$\textbf{157.29} \\
 \bottomrule
\end{tabular}
\label{Ta:example_fsi_macro}
\end{table}

\section{Conclusion}\label{S:conclusion}
We introduced a new abstraction for
solving multi-domain problems using finite element methods,
which allows for representing multi-domain problems in
the mixed, or multivariable, variational problem formalism.
This enables mixed domain problems to be represented
by first class Firedrake construct,
hence enabling full composition of
discretisation solver interface as well as
outer loop capabilities such as adjoint simulations and deflation.

We implemented our new abstraction in UFL and Firedrake;
we introduced two new classes in UFL,
\code{MeshSequence} and \code{CellSequence}, and
modified \code{MixedElement} class.

We validated our implementations
by performing a convergence study with 
a quad-triangle mixed-cell-type problem and
a hex-quad mixed-cell-type problem, and
by solving a fluid-structure interaction benchmark problem.

We have so far only considered multi-domain problems involving
conforming submeshes, but
our new abstraction and implementation in UFL
can readily be extended for
problems involving non-conforming meshes.
The details will be left for the near future work.

\section*{Acknowledgment}
This work was funded under
the Engineering and Physical
Sciences Research Council [grant numbers EP/R029423/1, EP/W029731/1, EP/W026066/1].

\section*{Code availability}
The version of Firedrake, along with all of the scripts,
required to reproduce the experiments presented in this paper has been archived on Zenodo~\cite{firedrake_zenodo}.

\ifarxiv

\else
\bibliographystyle{ACM-Reference-Format}
\bibliography{reference}
\fi
\end{document}